\documentclass[10pt,twocolumn,aps,prd,preprintnumbers,showpacs,superscriptaddress,nofootinbib,amsmath,amssymb,floats,floatfix,showkeys,notitlepage,longbibliography]{revtex4-2}
\usepackage[T1]{fontenc}
\usepackage[utf8]{inputenc}
\usepackage{newtxtext}
\usepackage{newtxmath}
\usepackage{amsmath,amssymb,amsfonts}

\usepackage{xcolor,graphicx,float}
\usepackage{orcidlink}
\usepackage{mdframed}
\usepackage{hyperref} 

\begin{document}

\title{Spinning particle dynamics, epicyclic frequencies, and transient QPO signatures in Schwarzschild spacetime}

\author{Uktamjon Uktamov\orcidlink{0009-0003-0423-2474}} 
\email[Corresponding Author:]{uktam.uktamov11@gmail.com}
\affiliation{School of Physics, Harbin Institute of Technology, Harbin 150001, People’s Republic of China.}
\affiliation{Tashkent University of Applied Sciences, Gavhar Str. 1, Tashkent 100149, Uzbekistan.}
\affiliation{Institute for Advanced Studies, New Uzbekistan University,
Movarounnahr str. 1, Tashkent 100000, Uzbekistan
.}
\affiliation{Tashkent State Technical University, Tashkent 100095, Uzbekistan}

\author{Ali \"Ovg\"un \orcidlink{0000-0002-9889-342X}}
\email{ali.ovgun@emu.edu.tr}
\affiliation{Physics Department, Faculty of Arts and Sciences, Eastern Mediterranean University, Famagusta, 99628 North Cyprus via Mersin 10, Turkiye.}

\author{Reggie C. Pantig \orcidlink{0000-0002-3101-8591}} 
\email{rcpantig@mapua.edu.ph}
\affiliation{Physics Department, School of Foundational Studies and Education, Map\'ua University, 658 Muralla St., Intramuros, Manila 1002, Philippines.}
\author{Bobomurat Ahmedov \orcidlink{0000-0002-1232-610X}}
\email[Corresponding Author:]{ahmedov@astrin.uz}
\affiliation{School of Physics, Harbin Institute of Technology, Harbin 150001, People’s Republic of China.}
\affiliation{Institute for Advanced Studies, New Uzbekistan University,
Movarounnahr str. 1, Tashkent 100000, Uzbekistan
.}
\affiliation{Institute of Theoretical Physics, National University of Uzbekistan, Tashkent 100174, Uzbekistan.}

\date{\today}

\begin{abstract}
    We study the motion of spinning test particles in Schwarzschild spacetime within the Mathisson--Papapetrou--Dixon pole--dipole approximation, imposing the Tulczyjew--Dixon spin supplementary condition. Restricting to equatorial orbits with the particle spin aligned with the orbital angular momentum, and retaining terms through linear order in the specific spin $s$, we derive the spin-corrected radial potential, circular-orbit conditions, bound periodic trajectories, epicyclic frequencies, and Lyapunov exponents of unstable circular orbits. The spin--curvature coupling shifts the circular-orbit energy and angular momentum and moves the innermost stable circular orbit to $r_{\rm ISCO}=6M-2\sqrt{2/3}\,s+\mathcal{O}(s^2)$ in the sign convention adopted here. We construct bound periodic orbits using the Levin--Perez-Giz zoom--whirl taxonomy and show how the particle spin deforms the corresponding energy--angular-momentum map. We then obtain the coordinate-time azimuthal and radial epicyclic frequencies and use them as kinematical inputs for relativistic-precession and resonance prescriptions for quasi-periodic oscillations. Finally, we relate the Lyapunov exponent of unstable circular orbits to the local separatrix structure governing near-homoclinic zoom--whirl motion. The resulting formulation provides a compact analytic connection between linear-in-spin MPD dynamics, periodic-orbit taxonomy, epicyclic-frequency shifts, and transient strong-field phenomenology in a nonrotating black-hole background. Also, we study the gravitational waveforms from the periodic orbits of a massive spinning particle around a  black hole, presenting those associated with extreme mass-ratio inspirals involving a stellar-mass compact spinning object orbiting a supermassive black hole.
\end{abstract}


\maketitle
\section{Introduction} \label{sec1}

Black holes provide one of the cleanest arenas in which to test the nonlinear and strong-field predictions of general relativity \cite{Berti:2015itd,Cardoso:2019rvt,Vagnozzi:2022moj,Battista:2023iyu,Battista:2026nsx}.  The orbital motion of compact objects around black holes encodes information about the background geometry, the multipolar structure of the source, and the internal degrees of freedom of the orbiting body.  In the idealized point-particle limit, the motion is geodesic; however, realistic compact probes possess spin, and their trajectories are consequently affected by spin--curvature coupling.  The appropriate covariant description is supplied by the Mathisson--Papapetrou--Dixon (MPD) equations, supplemented by a spin supplementary condition that fixes the representative worldline of the extended body \cite{Mathisson:1937NM,Papapetrou:1951,Tulczyjew:1959,Dixon:1964}.  In the pole--dipole approximation, these equations provide a controlled framework for studying the leading finite-size correction to geodesic motion. The applicability of the pole--dipole approximation requires the characteristic size of the body to be much smaller than the local curvature scale, and the dimensionless spin measure $|s|/M$ must remain perturbative when the central mass $M$ sets the background scale.  For a compact secondary of mass $\mu$ and dimensionless spin $\chi$, one has $s=S/\mu\simeq \chi \mu$, so that $s/M\simeq \chi\,\mu/M$ is naturally small in the extreme-mass-ratio regime.  The Tulczyjew--Dixon condition used below selects the centroid measured in the zero-momentum frame and is especially convenient because the mass and spin magnitude are conserved along the MPD trajectory.  These restrictions should be kept in mind when interpreting large-spin curves: they are useful for displaying trends, whereas the controlled expansion is the small-$s/M$ regime \cite{Costa:2014nta,Semerak:1999qc,Kyrian:2007zz}.

The dynamics of spinning particles is particularly relevant for relativistic astrophysics.  In extreme-mass-ratio inspirals, a stellar-mass compact object moves for a long time in the strong-field region of a massive black hole, and even small corrections to the orbital frequencies can accumulate into observable phase shifts.  Similarly, in X-ray binaries and galactic-center systems, the characteristic orbital, radial, and vertical epicyclic frequencies are often used to model quasi-periodic oscillations (QPOs) and to infer black-hole parameters from timing data \cite{Stella:1998lense,Stella:1999prl,Kluzniak:2001ar,Abramowicz:2001bi}.  The inclusion of particle spin therefore provides a natural extension of geodesic frequency models and can clarify how finite-size effects modify strong-field observables. At the same time, QPO identifications are intrinsically model dependent: the same observed pair of frequencies may be interpreted through relativistic precession, diskoseismic, resonance, or geometric-precession mechanisms.  A particle-frequency calculation should therefore be viewed as a necessary kinematical input rather than a complete emission model.  In the present setting, this distinction is useful because the MPD equations determine how the orbital frequency map is shifted by the spin of the probe, while the conversion of this map into an observed X-ray timing signal requires additional assumptions about the radiating plasma and its modulation mechanism \cite{Remillard:2006fc,Abramowicz:2011xu,Ingram:2019mna}.

The Schwarzschild spacetime is the simplest black-hole background, but it remains a useful theoretical laboratory.  It isolates the spin--curvature interaction of the test body from frame-dragging effects of the central object and allows several aspects of the dynamics to be treated analytically.  The geodesic orbital structure of Schwarzschild black holes is well understood \cite{Chandrasekhar:1983}; nevertheless, spinning-particle motion is richer because the momentum and velocity need not be parallel beyond leading order, the effective radial potential acquires spin-dependent terms, and the innermost stable circular orbit (ISCO), bound-orbit taxonomy, and epicyclic frequency map are shifted by the spin of the probe.  Recent analytic progress has shown that the leading-order-in-spin MPD dynamics in spherically symmetric spacetimes can be solved in closed form, including representations in terms of elliptic functions \cite{Witzany:2023bmq}.  This motivates a systematic study connecting analytic spinning-particle orbits with frequency-based phenomenology.

The effect of spin on black-hole orbits has been studied from several complementary viewpoints.  Earlier analyses showed that spin--curvature coupling can produce substantial departures from geodesic motion and, outside the strictly perturbative regime, even chaotic behavior in Schwarzschild spacetime.  The ISCO of a spinning particle in Schwarzschild and Kerr backgrounds has also been treated analytically in the small-spin limit.  The present work is therefore not intended merely as a rederivation of the spin-shifted ISCO; rather, its purpose is to connect the linear-in-spin circular-orbit corrections, periodic bound trajectories, epicyclic-frequency shifts, and local instability exponents within one Schwarzschild MPD framework \cite{Suzuki:1996gm,Jefremov:2015gza,Kyrian:2007zz}.

A complementary way to characterize strong-field bound motion is through periodic orbits.  In the Levin--Perez-Giz taxonomy, a periodic orbit is labeled by rational data that specify the number of zooms, whirls, and leaves traced by the trajectory \cite{Levin:2008mq}.  Such orbits are not merely mathematical curiosities: they organize the phase space of generic bound motion and are closely connected to separatrix behavior, zoom--whirl dynamics, and the harmonic content of gravitational radiation.  When the orbiting body carries spin, the spin--curvature force shifts the energy--angular-momentum map associated with a given periodic orbit, thereby modifying both the orbital morphology and the emitted waveform.  This provides a direct bridge between the analytic MPD dynamics and gravitational-wave modeling based on kludge or semi-analytic prescriptions \cite{Alloqulov:2026regularBH,Yang:2025qcbh}. For waveform applications, periodic orbits are also valuable because they isolate the harmonic structure associated with repeated radial librations and whirl phases.  Numerical-kludge waveforms provide a computationally inexpensive way to translate such strong-field trajectories into approximate gravitational-wave polarizations, although they do not replace black-hole perturbation theory or self-force waveforms when high-precision phase accuracy is required \cite{Babak:2006uv}.

In this work, we investigate spinning test-particle motion around a Schwarzschild black hole within the MPD pole--dipole approximation, imposing the Tulczyjew--Dixon spin supplementary condition and restricting to the aligned-spin sector.  We work consistently to linear order in the particle spin parameter, which is the regime in which the pole--dipole approximation is physically controlled.  Our aim is to treat these ingredients within a single perturbative MPD framework and to identify how the linear spin--curvature coupling modifies the orbital quantities entering QPO and zoom--whirl analyses.

The main results may be summarized as follows.  First, we derive the spin-corrected radial potential and obtain the circular-orbit energy and angular momentum to linear order in the particle spin.  We show how the spin--curvature interaction shifts the ISCO relative to the geodesic Schwarzschild value.  Second, we construct analytic bound trajectories in terms of elliptic functions and apply the Levin--Perez-Giz taxonomy to spinning-particle periodic orbits.  Third, we compute the coordinate-time azimuthal and radial epicyclic frequencies and use them to discuss relativistic-precession and resonance-based QPO models.  Fourth, we analyze the Lyapunov exponent of unstable circular orbits and discuss its relevance for near-separatrix transient features.  Finally, we illustrate how periodic spinning-particle orbits can be used as input for numerical-kludge gravitational-wave signals.

The paper is organized as follows.  In Sec. \ref{sec2} we review the aligned-spin MPD equations in Schwarzschild spacetime and derive the effective potential, circular-orbit conditions, and ISCO quantities.  In Sec. \ref{sec3} we construct analytic periodic orbits and discuss their classification through the zoom--whirl taxonomy.  In Sec. \ref{sec4} we derive the epicyclic frequencies of spinning particles and connect them to QPO phenomenology.  In Sec. \ref{sec5} we study the Lyapunov instability of unstable circular orbits and its implications for transient QPO-like signals.  In Sec. \ref{sec6} we discuss unbound trajectories and periastron-precession constraints.  In Sec. \ref{sec7} we present numerical-kludge gravitational-wave signals generated by representative periodic orbits.  We conclude in Sec. \ref{sec8}.  Throughout the paper we use geometrized units $G=c=1$.

\section{Motion of the spinning particle in the vicinity of the Schwarzschild BH} \label{sec2}
The space-time metric of the Schwarzschild BH can be expressed as:
\begin{eqnarray}\label{eq.the line}
    ds^2&=&-f(r)dt^2+\frac{dr^2}{f(r)}+r^2d\Omega^2\,,\\\nonumber
    f(r)&=&1-\frac{2M}{r}\,,
\end{eqnarray}
in which $d\Omega^2=d\theta^2+\sin^2{\theta}d\phi^2$. 

The motion of a spinning test particle is described by the Mathisson-Papapetrou-Dixon (MPD) equation, which can be expressed as (\cite{Mathisson:1937NM,Papapetrou:1951,Dixon:1964}):
\begin{eqnarray}\label{eq. MPD}
    \frac{Dp^\mu}{d\lambda}&=&-\frac{1}{2}R^\mu_{\nu\kappa\gamma}\Dot{x}^\nu S^{\kappa\gamma}\,,\\\nonumber
    \frac{DS^{\mu\nu}}{d\lambda}&=&p^\mu\Dot{x}^\nu-p^\nu\Dot{x}^\mu\,,
\end{eqnarray}
where  $\frac{D}{d\lambda}$ denotes the covariant derivative along the path, where $\lambda$ serves as the affine parameter, so that $\frac{D}{d\lambda}=u^\mu\nabla_\mu$ and the quantity $R^\mu_{\nu\kappa\gamma}$ represents the Riemann tensor.
 Additionally, the particle's dynamical four momentum and kinematical four velocity are represented by $p^\mu$ and $u^\mu$, while $S^{\mu\nu}$ denotes the antisymmetric spin tensor, satisfying $S^{\mu\nu}=-S^{\nu\mu}$. 

The Tulczyjew--Dixon spin supplementary condition $S^{\mu\nu}p_\nu=0$ enables to obtain expression for four momentum of the particle with mass $m$ as (\cite{Witzany:2023bmq}) $p^\mu=m\Dot{x}^\mu+\mathcal{O}(S^2)$ and expression for the spin vector is $S^{\mu\nu}=\frac{1}{2}m\epsilon^{\mu\nu\kappa\gamma}\Dot{x}_\kappa s_\gamma$ with the specific spin vector $s^\gamma$ and Levi-Civita pseudo-tensor $\epsilon^{\mu\nu\kappa\gamma}$.  Hence, MPD equation (\ref{eq. MPD}) is simplified as:
 \begin{eqnarray}\label{eq.MPD2}
   \frac{D^2x^\mu}{d\lambda^2}&=&-\frac{1}{4}R^\mu_{\nu\gamma\delta}\epsilon^{\gamma\delta}_{\kappa\alpha}\Dot{x}^\nu\Dot{x}^\kappa s^\alpha\,,\\\nonumber
   \frac{Ds^\alpha}{d\lambda}&=&0\,.
 \end{eqnarray}
 Then the equation of  motion for particle with fully aligned spin vector $s$ can be expressed as (\cite{Witzany:2023bmq}):
 \begin{subequations}\label{eq.equation of motion}
    \begin{align}
        &\Dot{r}^2=\mathcal{E}^2-f(r)\left[1+\frac{l^2}{r^2}\right]+2\frac{s\mathcal{E}l}{r^2}\left[1-\frac{3M}{r}\right]=\\\nonumber
        &=\left(\mathcal{E}-V_{eff}^-\right)\left(\mathcal{E}-V_{eff}^+\right)\,,\\
        &\Dot{t}=\frac{\mathcal{E}}{f(r)}+s\frac{lM}{r^3f(r)}\,,\\
        &\Dot{\phi}=\frac{l}{r^2}+s\frac{\mathcal{E}}{r^2}\,,
    \end{align} 
    \end{subequations}
where $\mathcal{E}$ and $l$ are specific energy and specific angular momentum of the spinning particle, respectively. Hence, effective potential $V_{eff}$ for the spinning particle can be expressed as(\cite{Khan:2024jez}):
\begin{eqnarray}\label{eq.Veff}
    V_{eff}^\pm=-\frac{ls}{r^2}\left[1-\frac{3M}{r}\right]\pm\sqrt{f(r)\left[1+\frac{l^2}{r^2}\right]+\frac{l^2s^2}{r^4}\left[1-\frac{3M}{r}\right]^2}\,.
\end{eqnarray}   

\begin{figure*}[ht!]
\includegraphics[width=0.45\textwidth]{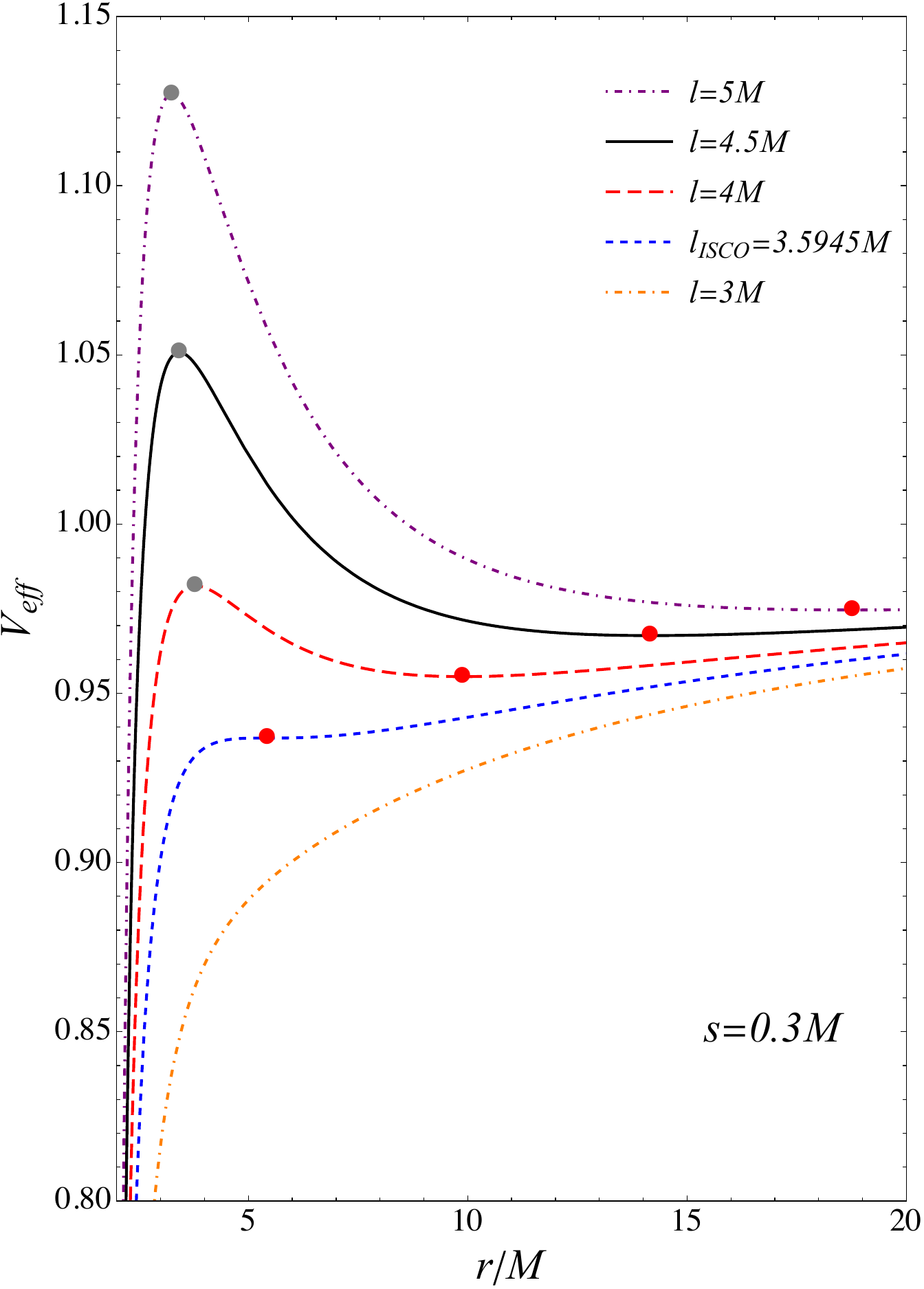}
\includegraphics[width=0.45\textwidth]{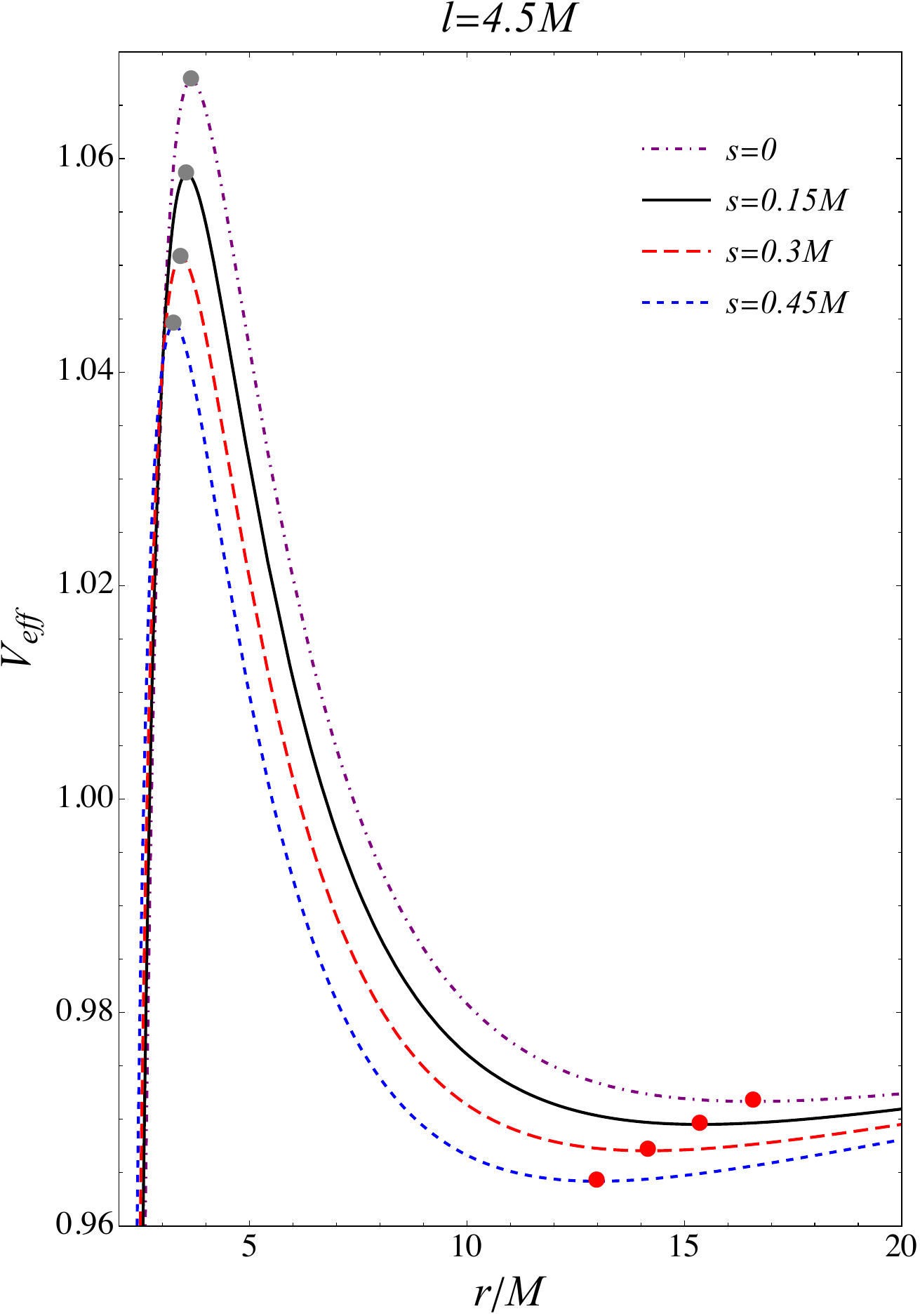}
\caption{The effective potential is shown as a function of the radial coordinate, plotted for constant values of the specific angular momentum $l$ (left panel) and the spin parameter $s$ (right panel). In the effective potential, the minima (red points) represent stable circular orbits at $r_{\text{min}}$, whereas the maxima (gray points) correspond to unstable circular orbits at $r_{\text{max}}$. \label{Fig.effective}}
\end{figure*}
  
Then we have plotted radial dependence of the effective potential $V_{eff}$ for different cases in Fig.(\ref{Fig.effective}). One can see from plots in Fig.(\ref{Fig.effective}) that effective potential $V_{eff}$ has two extreme points which correspond to the stable circular orbits (red points) and unstable circular orbits (gray points) (\cite{Uktamov:2024magdip}). It is clear from right panel of Fig.(\ref{Fig.effective}) that increasing the value of the spin parameter $s$ causes shrinking the radii of the stable and unstable circular orbits, however we can see from right panel of the Fig.(\ref{Fig.effective}) that increasing the value of the specific angular momentum $l$ yields growing the value of the radii of the stable and unstable circular orbits which traditional effect of $l$ on the orbits of the test particles.

Also, one can conclude from Fig.(\ref{Fig.effective}) that after some small values of the specific angular momentum $l_{ISCO}>l$ there is not any circular orbits of the test particle. These extreme values of the angular momentum $l_{ISCO}$ and corresponding specific energy $\mathcal{E}_{ISCO}$ are called energy and angular momentum at the innermost stable circular orbits (ISCO) and ISCO parameters can be found using conditions (\cite{Uktamov:2025dmhalo,Uktamov:2025stringy,Uktamov:2025s2mcmc,Uktamov:2024selfdual}):
\begin{eqnarray}\label{eq.ISCO}
V_{eff}\vert_{r=r_{ISCO}}=\mathcal{E}_{ISCO}\,,\,\,\,\frac{\partial V_{eff}}{\partial r}=\frac{\partial^2V_{eff}}{\partial r^2}=0\vert_{r=r_{ISCO}}\,.
\end{eqnarray}

\begin{figure*}[t]
\includegraphics[width=0.3245\textwidth]{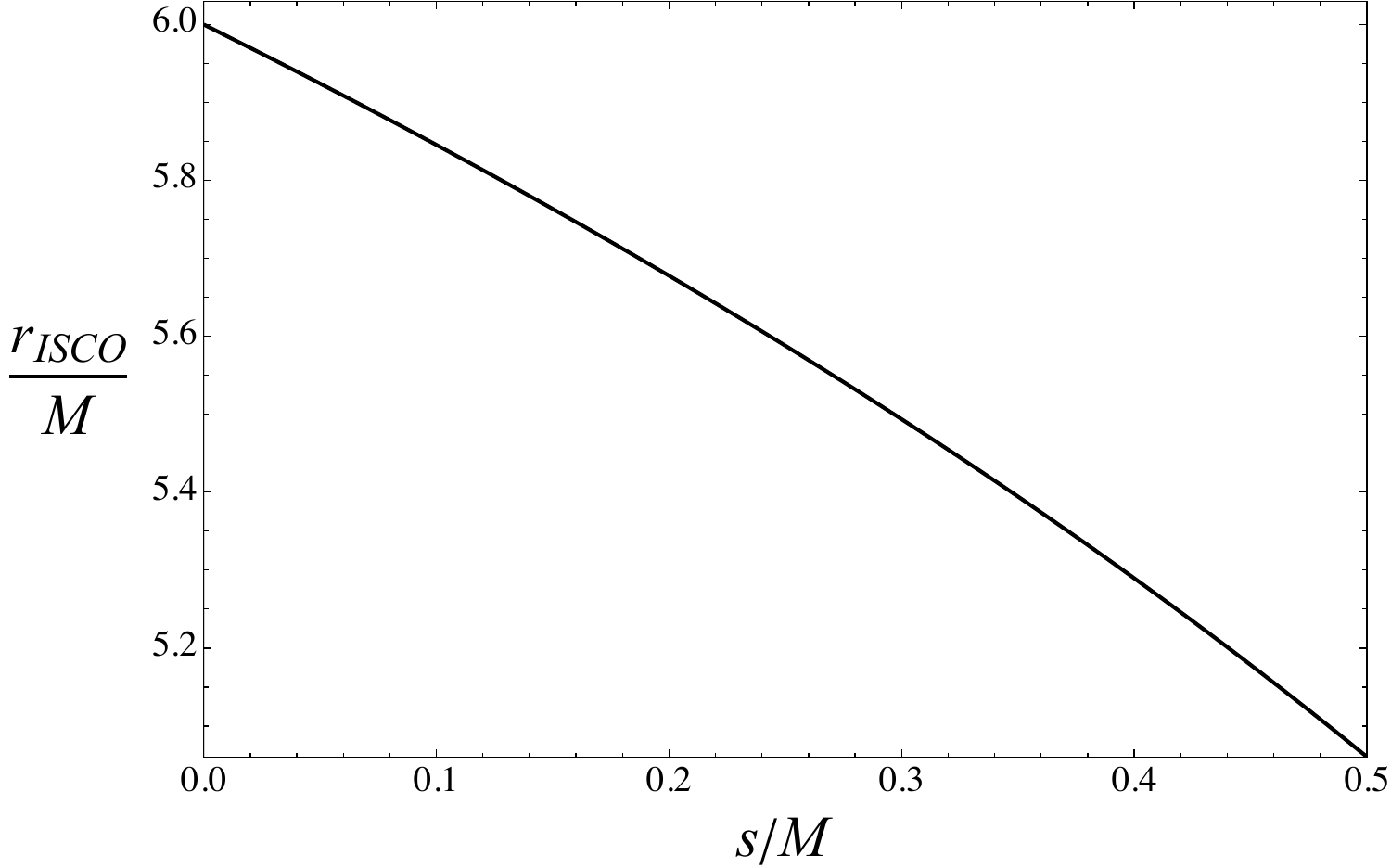}
\includegraphics[width=0.3245\textwidth]{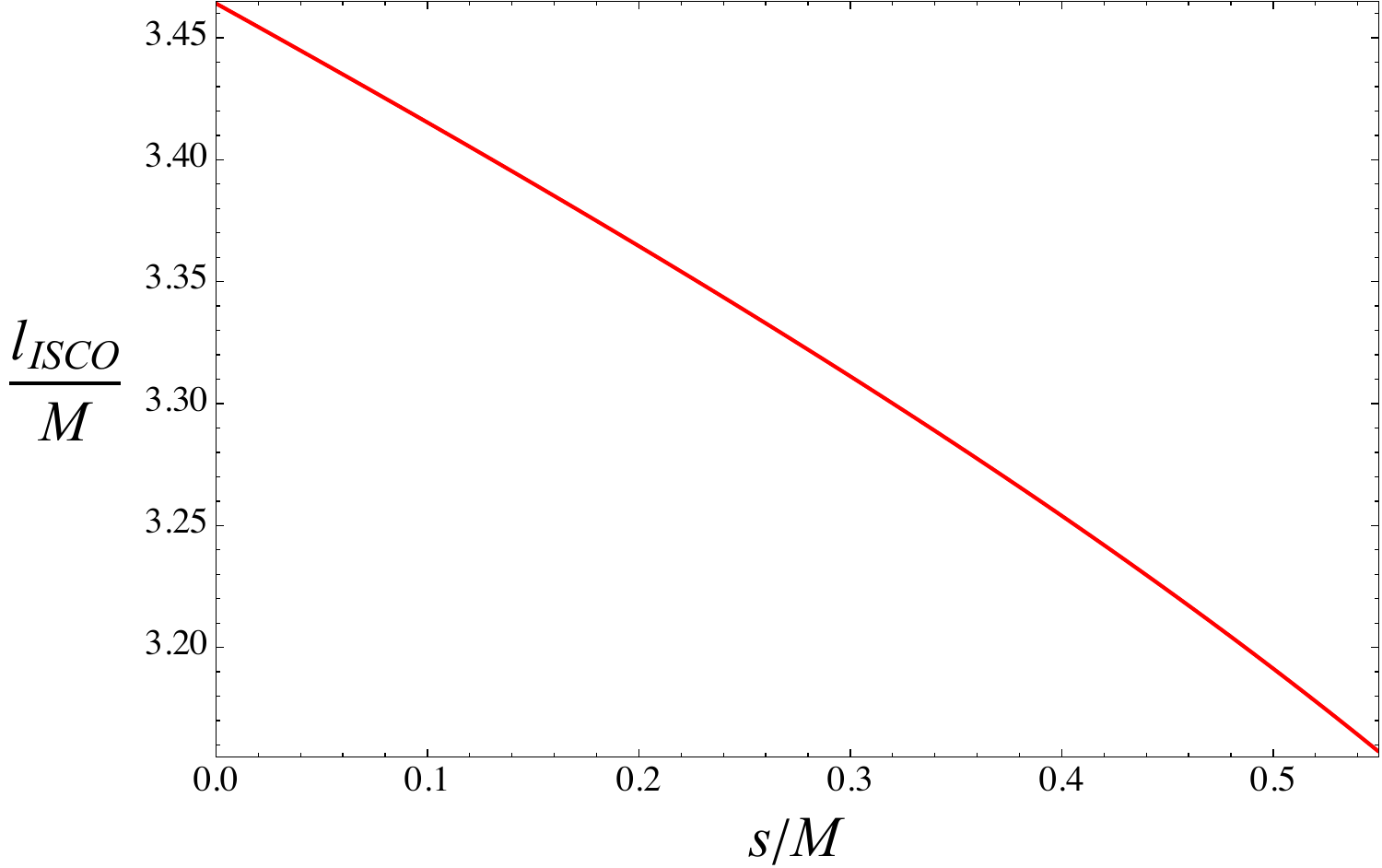}
\includegraphics[width=0.3245\textwidth]{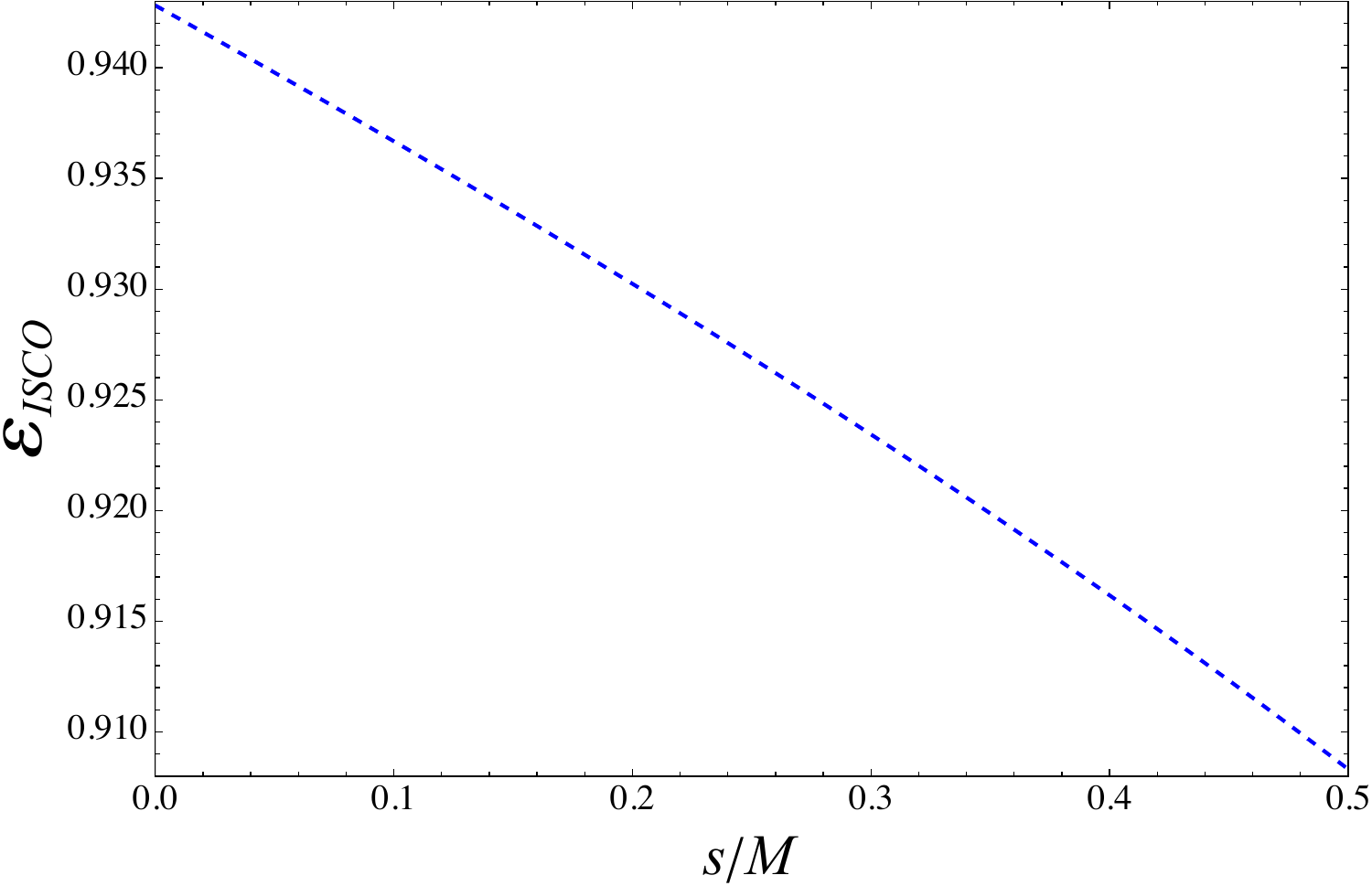}
\caption{
Plot showing the ISCO radius $r_{\text{ISCO}}$, specific angular momentum $l_{\text{ISCO}}$, and specific energy $\mathcal{E}_{\text{ISCO}}$ as functions of the spin parameter $s/M$.\label{Fig.ISCO}}
\end{figure*}

Then we have plotted dependence of the ISCO parameters $r_{ISCO}$, $l_{ISCO}$, $\mathcal{E}_{ISCO}$ on the spin parameter $s/M$ in Fig.(\ref{Fig.ISCO}). It is clear from plots in Fig.(\ref{Fig.ISCO}) that increasing the value of the spin parameter $s$ leads to decrease the value of the ISCO parameters $r_{ISCO}$, $l_{ISCO}$, $\mathcal{E}_{ISCO}$. 

The general equation of the motion at the equatorial plane for spinning particle (\ref{eq.equation of motion}) yields to the differential equation (\cite{UktamjonUktamov:2026dep,Ovgun:2025ehi,UktamjonUktamov:2025qts,Uktamjon:2024fjb}):

 \begin{widetext}  {\color{black}
\begin{eqnarray}\label{eq.dr}
    \frac{dr}{d\phi}
    =
    \pm\frac{r^2}{l+s\mathcal{E}}
    \sqrt{
    (\mathcal{E}^2-1)
    +\frac{2M}{r}
    -\frac{l(l-2s\mathcal{E})}{r^2}
    +\frac{2Ml(l-3s\mathcal{E})}{r^3}
    }\, .
\end{eqnarray}}
\end{widetext}

 {\color{black} Subsequently, the specific energy $\mathcal{E}$ and specific angular momentum $l$ of a spinning particle on a circular orbit of radius $r_0$ are obtained from $\mathcal{R}(r_0)=0,\qquad \mathcal{R}'(r_0)=0$,
where
\begin{equation}\label{eq.R}
\mathcal{R}(r)=
\mathcal{E}^2
-f(r)\left(1+\frac{l^2}{r^2}\right)
+\frac{2s\mathcal{E}l}{r^2}\left(1-\frac{3M}{r}\right).
\end{equation}
Solving these conditions perturbatively to first order in the spin parameter $s$, one obtains
\begin{subequations}\label{eq.circular_EL_corrected}
\begin{align}
\mathcal{E}_c
&=
\mathcal{E}_0
-\frac{sM^{3/2}}{2r_0(r_0-3M)^{3/2}}
+\mathcal{O}(s^2),
\\
l_c
&=
l_0
+
\frac{s(2r_0-9M)(r_0-2M)}
{2\sqrt{r_0}(r_0-3M)^{3/2}}
+\mathcal{O}(s^2),
\end{align}
\end{subequations}
where
\begin{equation}
\mathcal{E}_0=\frac{r_0-2M}{\sqrt{r_0(r_0-3M)}} ,
\qquad
l_0=r_0\sqrt{\frac{M}{r_0-3M}}
\end{equation}
are the standard Schwarzschild geodesic circular-orbit energy and angular momentum, also whole expressions for $l_c$ and $\mathcal{E}_c$ are given in Appendix (\ref{apdxA}).}

\begin{figure}
\includegraphics[width=0.45\textwidth]{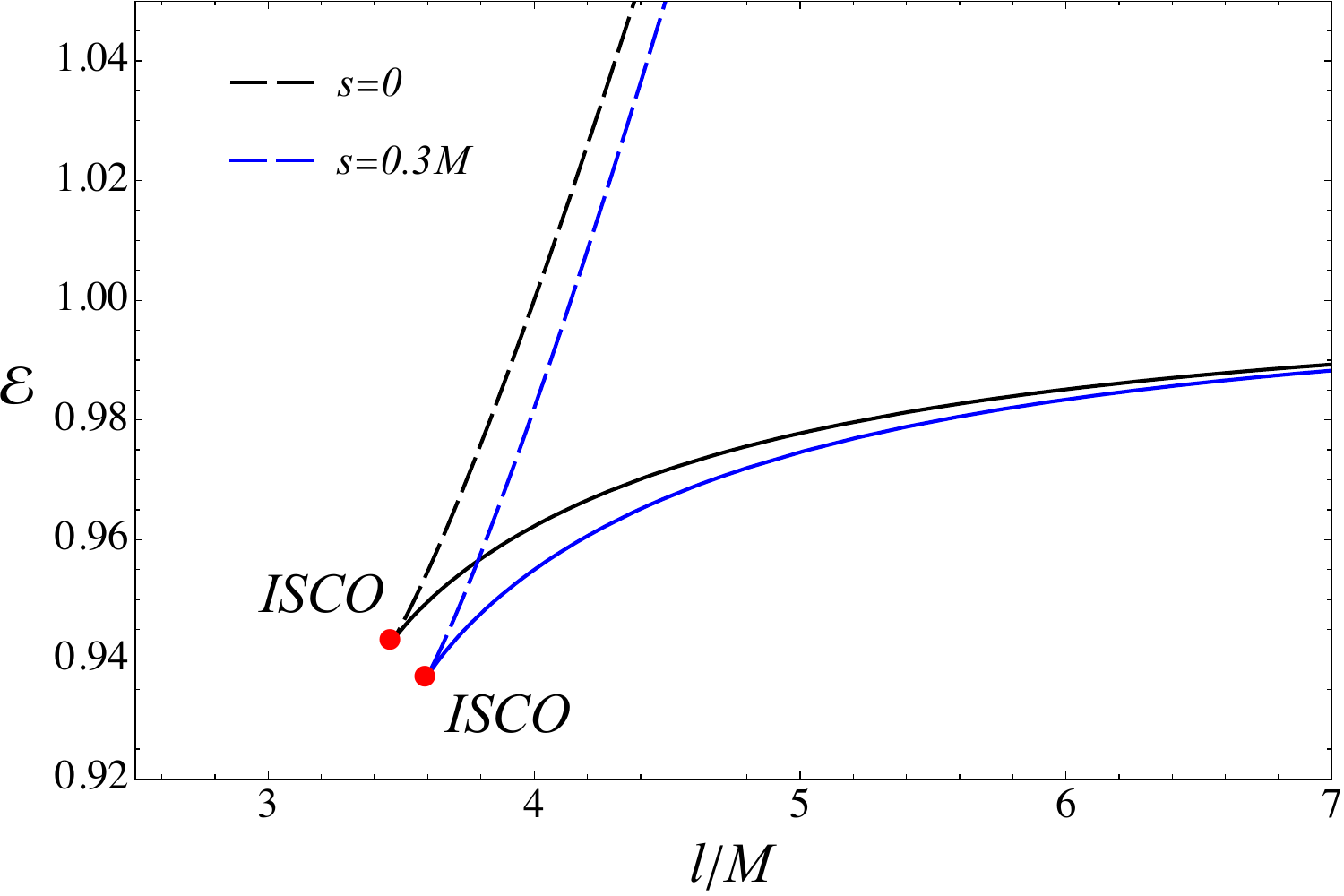}
\caption{The specific energy $\mathcal{E}$ and specific angular momentum $l$ for circular orbits are shown for various fixed values of the spin parameter $s$. Stable circular orbits are represented by solid lines, while dashed lines correspond to the unstable circular orbits.}
\label{Fig. space}
\end{figure}

\subsection{Analytical solution for periodic orbits of the spinning particles}
 To obtain analytic solution for the equation of motion of the spinning particle  Eq.(\ref{eq.dr}) should be rewritten as:
  \begin{eqnarray}\label{eq.dr1}
      \frac{dr}{d\phi}=\pm\sqrt{P(r)}\,,
  \end{eqnarray}
  in which
  \begin{widetext}
  \begin{eqnarray}
      P(r)=-\frac{r}{(l+s\mathcal{E})^2}\left[(1-\mathcal{E}^2)r^3-2Mr^2+lr(l-2s\mathcal{E})-2lM(l-3s\mathcal{E})\right]\,.
  \end{eqnarray}
  \end{widetext}
  Then we can conclude spinning particles may make periodic orbits (between apoapsis $r_1$ and periapsis $r_2$) if $\mathcal{E}<1$ so for the spinning particles making bound orbits we have(\cite{Uktamov:2024magdip,Alloqulov:2026modmax}):
  \begin{eqnarray}\label{eq.phi}
      \phi=\frac{l+s\mathcal{E}}{\sqrt{1-\mathcal{E}^2}}\int_{r_2}^r\frac{dr}{\sqrt{r(r_1-r)(r-r_2)(r-r_3)}}\,,
  \end{eqnarray}
 where the phase convention has been chosen such that the radial motion starts at periapsis, $\phi=0$ and $r=r_2$. A different starting point, for example apoapsis $r=r_1$, is obtained by shifting the elliptic-function argument by a half radial period. Then integrating Eq.(\ref{eq.phi}) yields:
  \begin{eqnarray}\label{eq.phi2}
      \phi(r)=\frac{2(l+s\mathcal{E})}{\sqrt{(1-\mathcal{E}^2)(r_1-r_3)r_2}}F(x_0,k_0)
  \end{eqnarray}
in which $F(x_0,k_0)$ is the incomplete elliptic integral of the first kind with modulus $k_0=\sqrt{\frac{(r_1-r_2)r_3}{(r_1-r_3)r_2}}$ and argument $x=\arcsin{\sqrt{\frac{(r_1-r_3)(r-r_2)}{(r_1-r_2)(r-r_3)}}}$. 

Then the radial equation of the spinning particle can be found using Eq.(\ref{eq.phi2}) as:
\begin{eqnarray}\label{eq.rphi}
    r(\phi)=\frac{r_2(r_1-r_3)-r_3(r_1-r_2)\text{sn}^2\left(\phi_0,k_0\right)}{r_1-r_3-(r_1-r_2)\text{sn}^2\left(\phi_0,k_0\right)}\,,
\end{eqnarray}
  here $sn(\phi_0,k_0)$ is the Jacobi elliptic sine function with argument $\phi_0=\frac{\sqrt{(1-\mathcal{E}^2)(r_1-r_3)r_2}}{2(l+s\mathcal{E})}\phi$.

Subsequently, the roots of the $P(r)=0$ in Eq.(\ref{eq.dr1}) can be expressed as (\cite{Chandrasekhar:1983,Uktamov:2024magdip}):
  \begin{subequations}\label{eq.roots}
      \begin{align}
          &r_1=\frac{\lambda}{1-e}\,,\\
          &r_2=\frac{\lambda}{1+e}\,,\\
          &r_3=\frac{2M}{1-\mathcal{E}^2}-\frac{2\lambda}{1-e^2}\,,
      \end{align}
  \end{subequations}
where $\lambda$ and $e$ ($0\leq e<1$ for bound orbits) are latus rectum and eccentricity, respectively. Hence, Eq.(\ref{eq.rphi}) can be expressed as:
  \begin{eqnarray}\label{eq.rphi2}
      r(\phi)=\frac{\lambda\left[\lambda\frac{3+e}{1+e}-2M\frac{1-e}{1-\mathcal{E}^2}\right]-2\lambda e\left[\frac{2M}{1-\mathcal{E}^2}-\frac{2\lambda}{1-e^2}\right]\text{sn}^2\left(\phi_0,k_0\right)}{\lambda(3+e)-2M\frac{1-e^2}{1-\mathcal{E}^2}-2\lambda e\text{sn}^2\left(\phi_0,k_0\right)}\,.
  \end{eqnarray}

By taking into account the oscillation of the radial coordinate between $r_1$ and $r_2$ over one orbital period, the corresponding change in the azimuthal angle can be obtained from Eq.(\ref{eq.phi2}):
\begin{eqnarray}\label{eq.phi3}
    \Delta\phi_r=2\phi(r_1)=\frac{4(l+s\mathcal{E})}{\sqrt{\lambda\left[\frac{\lambda(3+e)(1-\mathcal{E}^2)}{(1-e^2)(1+e)}-\frac{2M}{1+e}\right]}}K(k_0)\,,
\end{eqnarray}
here $K(k_0)$ represents the complete elliptic integral of the first kind.

Using the Levin--Perez-Giz taxonomy, Eq. \eqref{eq.phi3} can be written in terms of the rational zoom--whirl parameter $q=\omega+v/z$ as \cite{Levin:2008mq}:
\begin{eqnarray}
    q+1=\frac{2}{\pi}\frac{l_p+s\mathcal{E}_p}{\sqrt{\lambda\left[\frac{\lambda(3+e)(1-\mathcal{E}^2)}{(1-e^2)(1+e)}-\frac{2M}{1+e}\right]}}K(k_0)\,,
\end{eqnarray}
  where $q=\omega+\frac{v}{z}$ and $\omega$, $v$ and $z$ denotes whirl, vertex and zoom numbers, respectively. 

{\color{black}Additionally, for each choice of eccentricity $e$ and latus rectum $\lambda$, the corresponding specific energy and angular momentum may be obtained from the two turning-point conditions $P(r_1)=P(r_2)=0$, with
\begin{equation}
r_1=\frac{\lambda}{1-e},
\qquad
r_2=\frac{\lambda}{1+e}.
\end{equation}
Expanding to first order in the spin parameter gives
\begin{subequations}
\label{eq:Ep_lp_corrected}
\begin{align}
\mathcal{E}_p^2
=
\frac{(\lambda-2M)^2-4e^2M^2}
{\lambda\left[\lambda-M(3+e^2)\right]}
\nonumber\\
\quad
-
s\,
\frac{(1-e^2)^2M
\sqrt{M\lambda\left[\lambda^2-4M\lambda+4M^2(1-e^2)\right]}}
{\lambda^2\left[\lambda-M(3+e^2)\right]^2}
\\ \notag+\mathcal{O}(s^2),
\\
l_p^2
=
\frac{M\lambda^2}
{\lambda-M(3+e^2)}
\nonumber\\
\quad
+
s\,
\frac{\left[2\lambda-3M(3+e^2)\right]
\sqrt{M\lambda\left[\lambda^2-4M\lambda+4M^2(1-e^2)\right]}}
{\left[\lambda-M(3+e^2)\right]^2}
\\+\mathcal{O}(s^2) \notag.
\end{align}
\end{subequations}

 }
 Then we give some numerical values of the specific energy $\mathcal{E}$ and specific angular momentum $l$ corresponding to the chosen orbits in table (\ref{Table 1}).

\begin{table}[ht!]
    \centering
    \begin{tabular}{|c|c|c|c|c|}
     \hline
      $(z,\,\omega,\,v)$ & $s[M]$ & $\lambda[M]$& $\mathcal{E}_p$ & $l_p[M]$ \\
    \hline
      (2,0,1)   & 0  & 11.145 & 0.984504 & 4.06822 \\
      (2,1,1)   &    & 7.92327 & 0.97897 & 3.82839 \\
      (2,2,1)   &    & 7.64091 & 0.978327 & 3.82002 \\
      (3,0,1)   &    & 13.9918 & 0.987505 & 4.34877 \\
      (3,1,1)   &    & 8.06113 & 0.979274 & 3.8338 \\
      (3,2,1)   &    & 7.65755 & 0.979274 & 3.8204 \\
      (4,0,1)   &    & 16.9124 & 0.989591 & 4.64228 \\
      (4,1,1)   &    & 8.15226 & 0.979471 & 3.83779 \\
      (4,2,1)   &    & 7.66827 & 0.978391 & 3.82066 \\
           &    &  &  &\\
      (2,0,1)   & 0.3   & 13.4598 & 0.986983 & 4.54322 \\
      (2,1,1)   &    & 7.57145 & 0.977822 & 3.97938 \\
      (2,2,1)   &    & 7.04209 & 0.976421 & 3.95659 \\
      (3,0,1)   &    &  19.595 & 0.990964 & 5.1761 \\
      (3,1,1)   &    & 7.81183 & 0.978415 & 3.99368 \\
      (3,2,1)   &    & 7.07749 & 0.976519 & 3.95766 \\
      (4,0,1)   &    & 26.7747 & 0.993353 & 5.84796 \\
      (4,1,1)   &    & 7.96905 & 0.978789 & 4.00407 \\
      (4,2,1)   &    & 7.09962 & 0.97658 & 3.95836 \\
   \hline
    \end{tabular}
    \caption{{The specific energy $\mathcal{E}$ and specific angular momentum $l$ values corresponding to eccentricity $e=0.8$ and the different periodic orbit. Here to calculate $\mathcal{E}$ and $l$ we have used full expression given in Appendix (\ref{apdx2})}}
    \label{Table 1}
\end{table}

\begin{figure*}[t]
\includegraphics[width=0.34\textwidth]{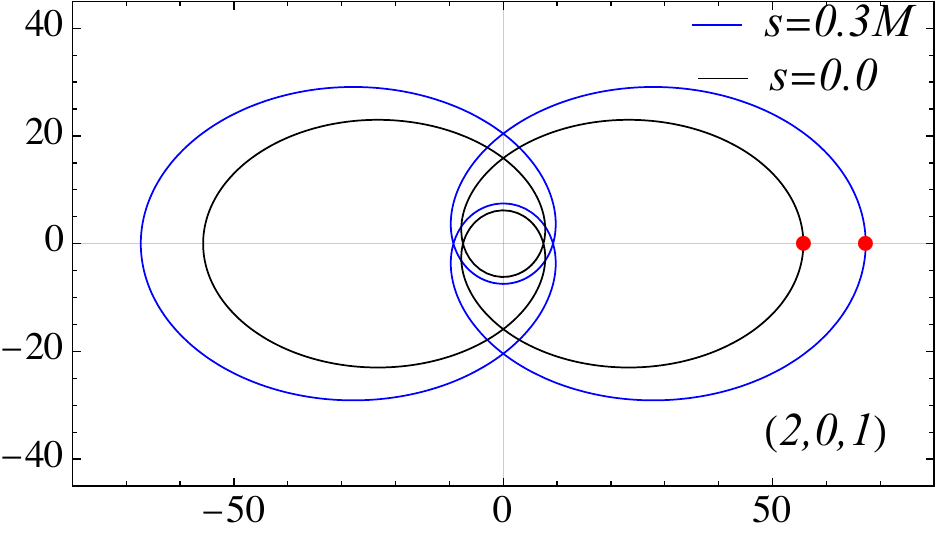}
\includegraphics[width=0.31\textwidth]{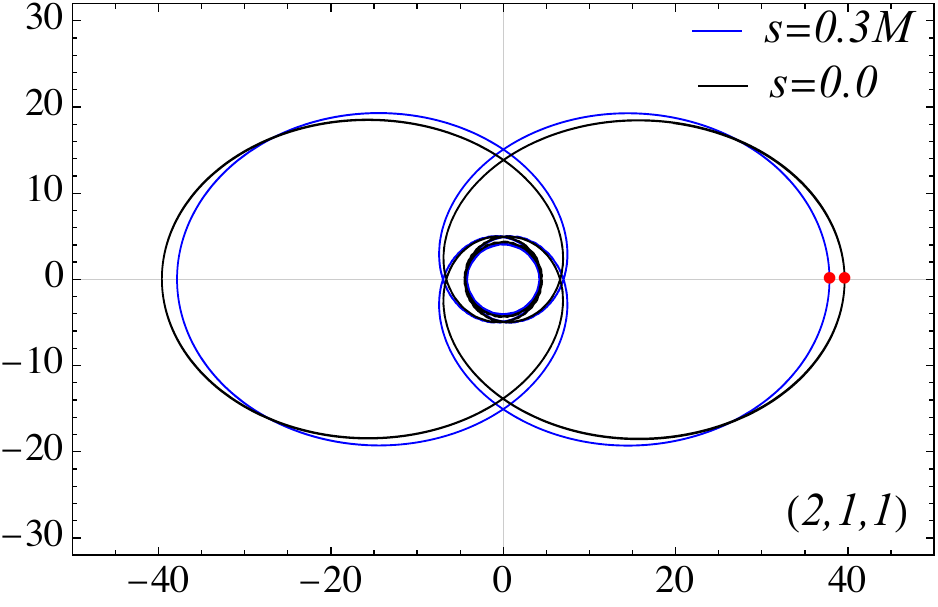}
\includegraphics[width=0.32\textwidth]{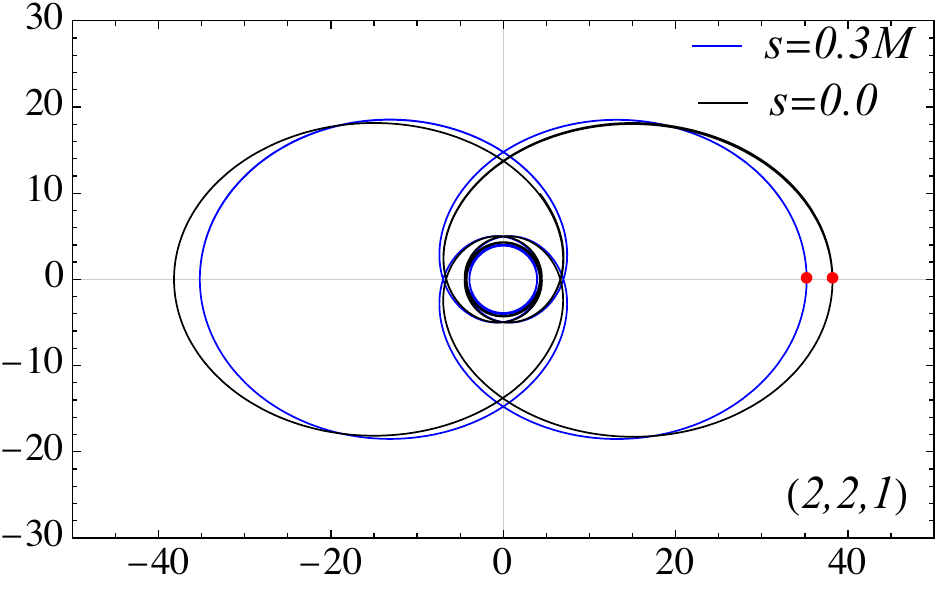}
\includegraphics[width=0.33\textwidth]{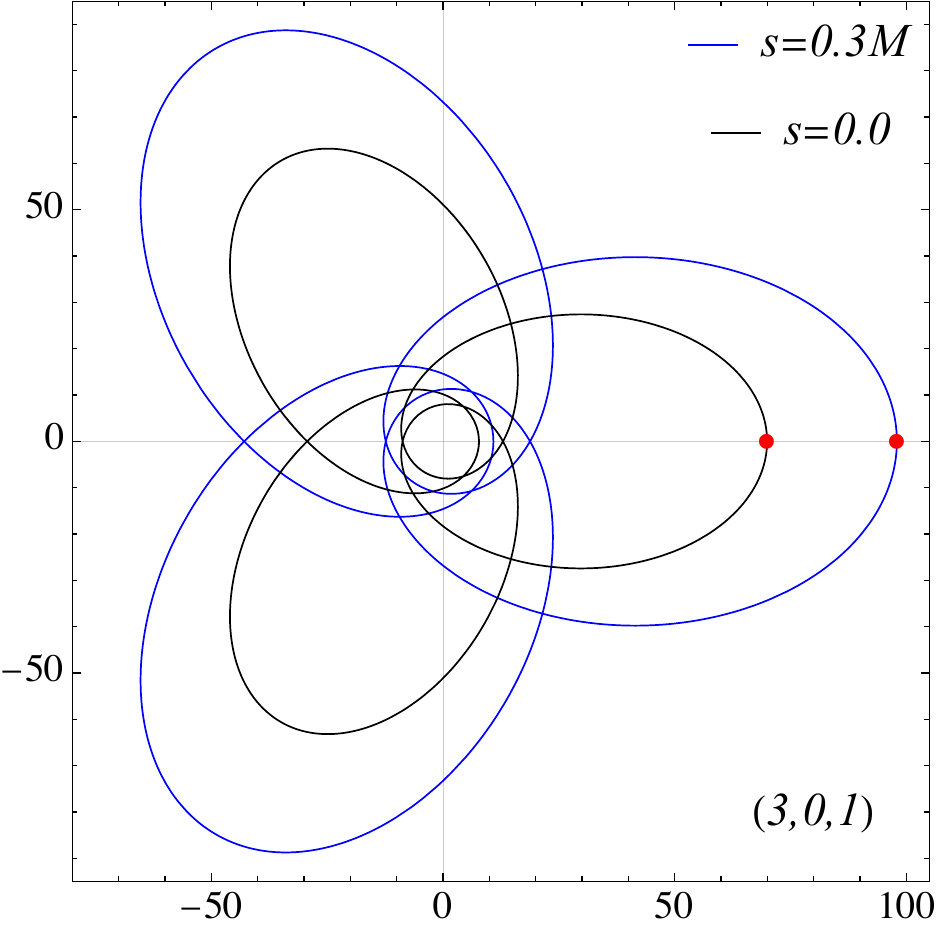}
\includegraphics[width=0.33\textwidth]{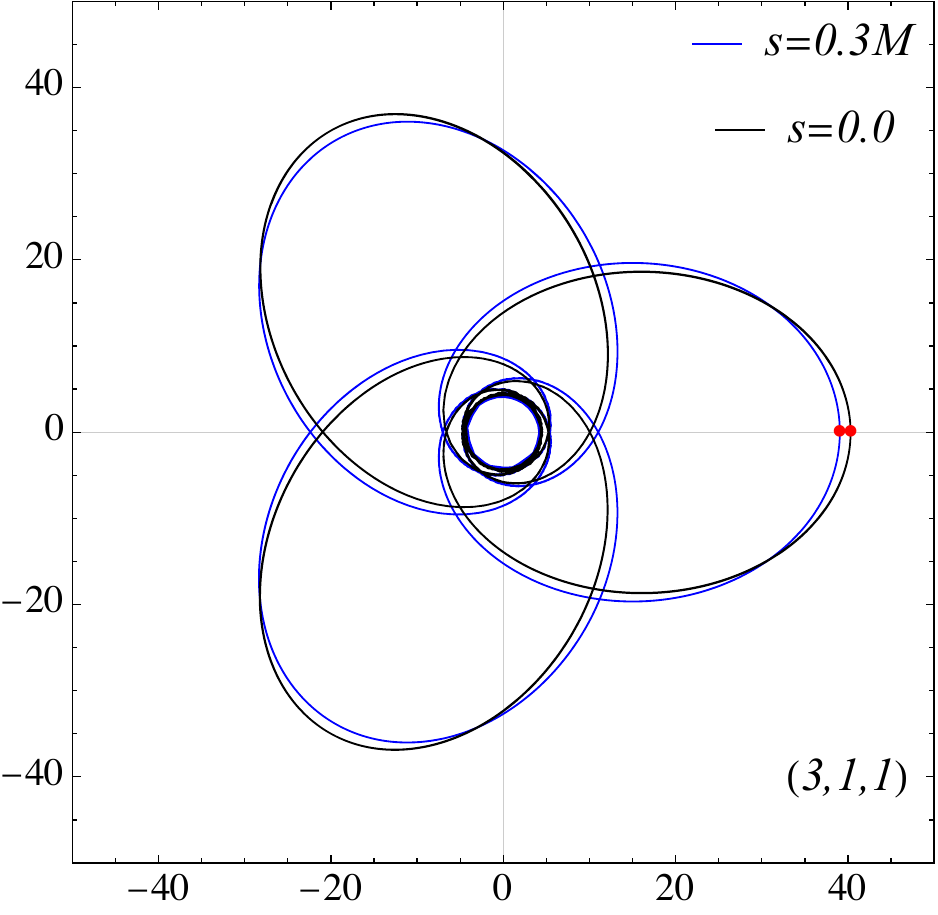}
\includegraphics[width=0.33\textwidth]{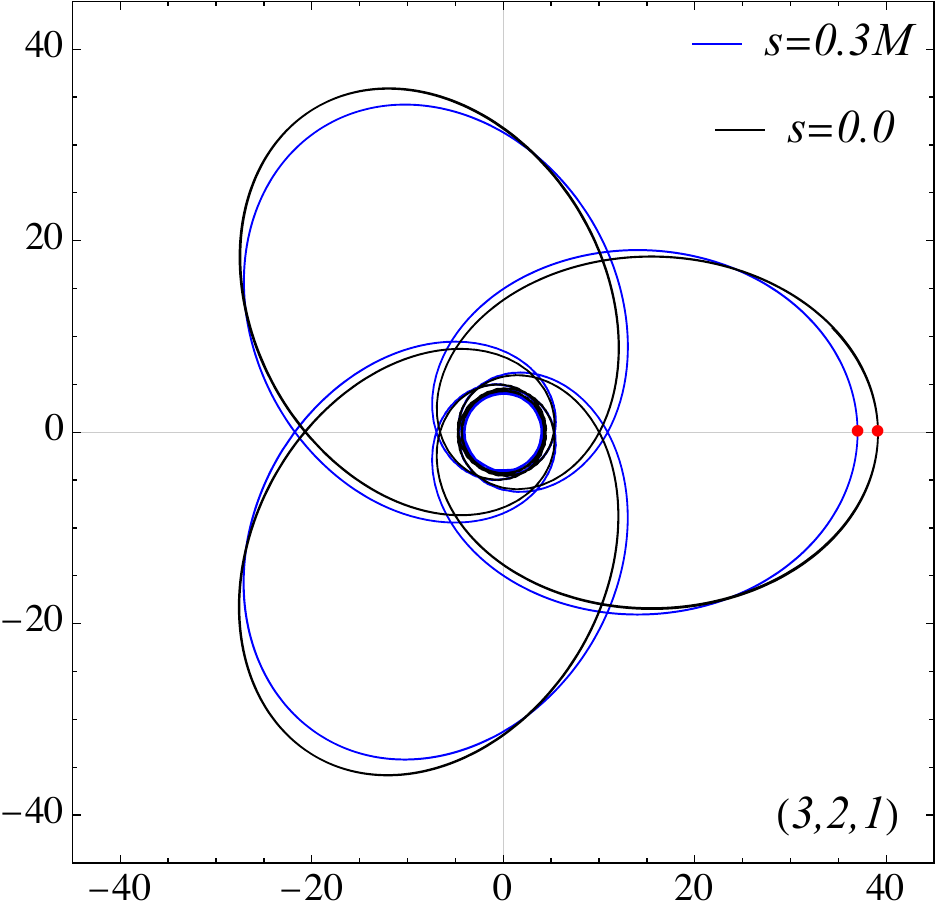}
\includegraphics[width=0.33\textwidth]{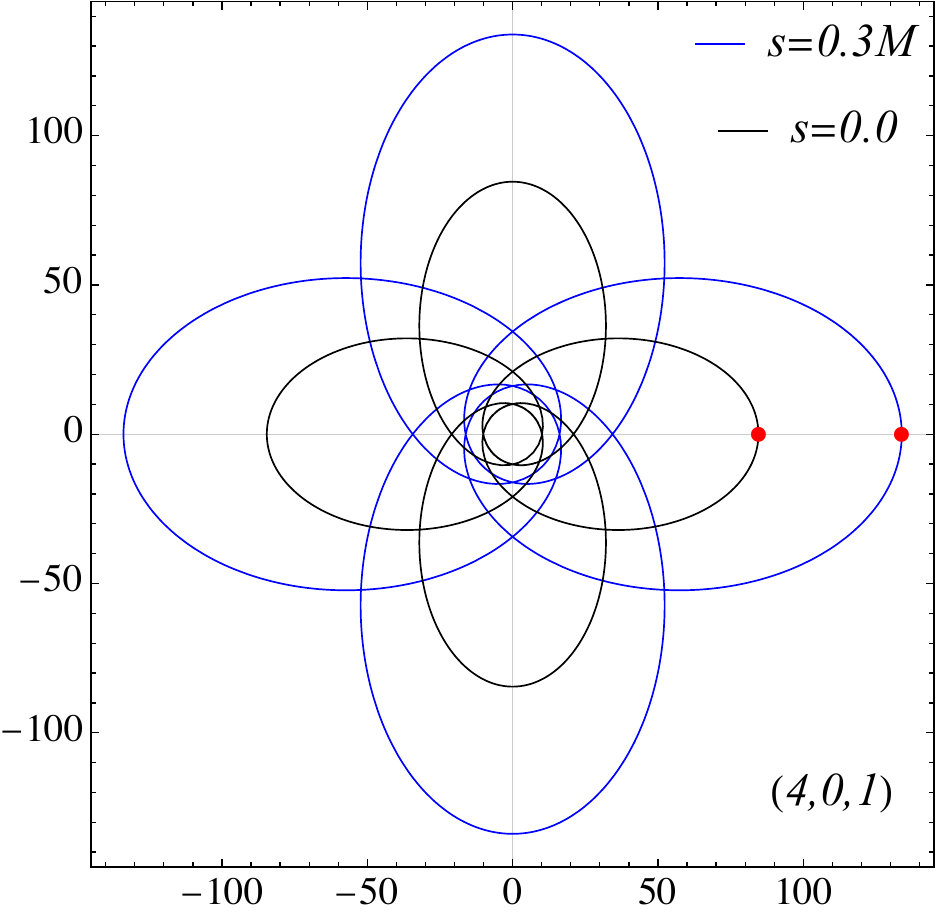}
\includegraphics[width=0.33\textwidth]{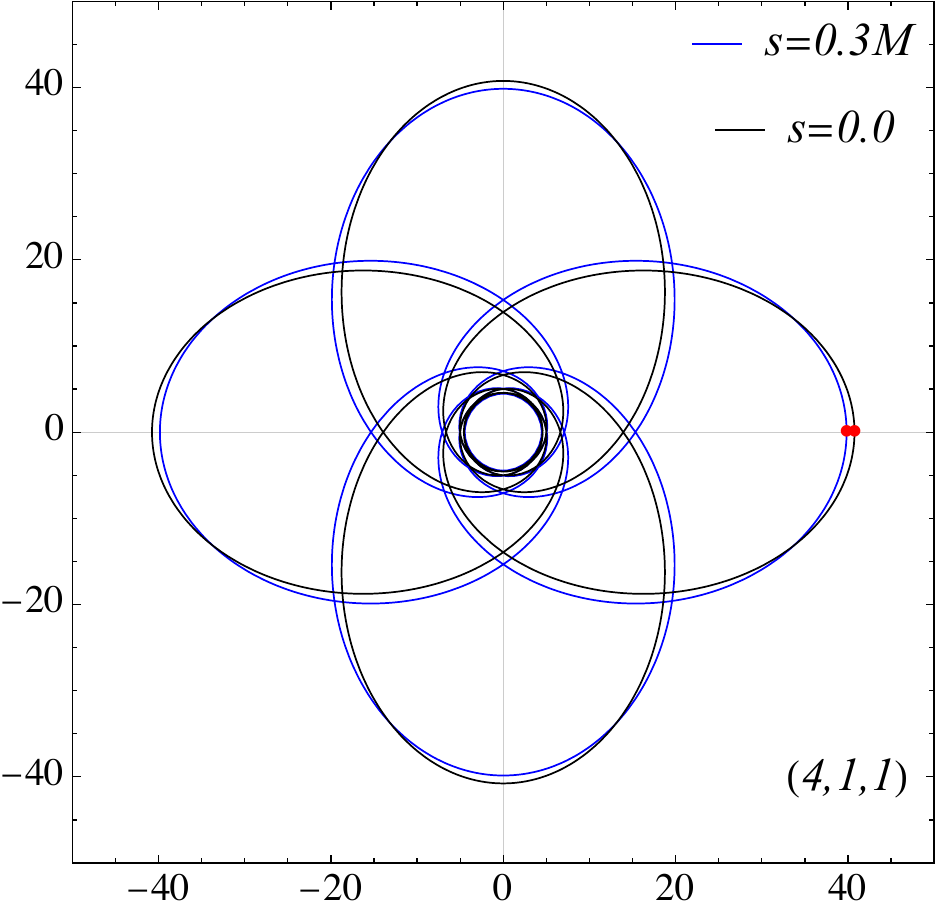}
\includegraphics[width=0.33\textwidth]{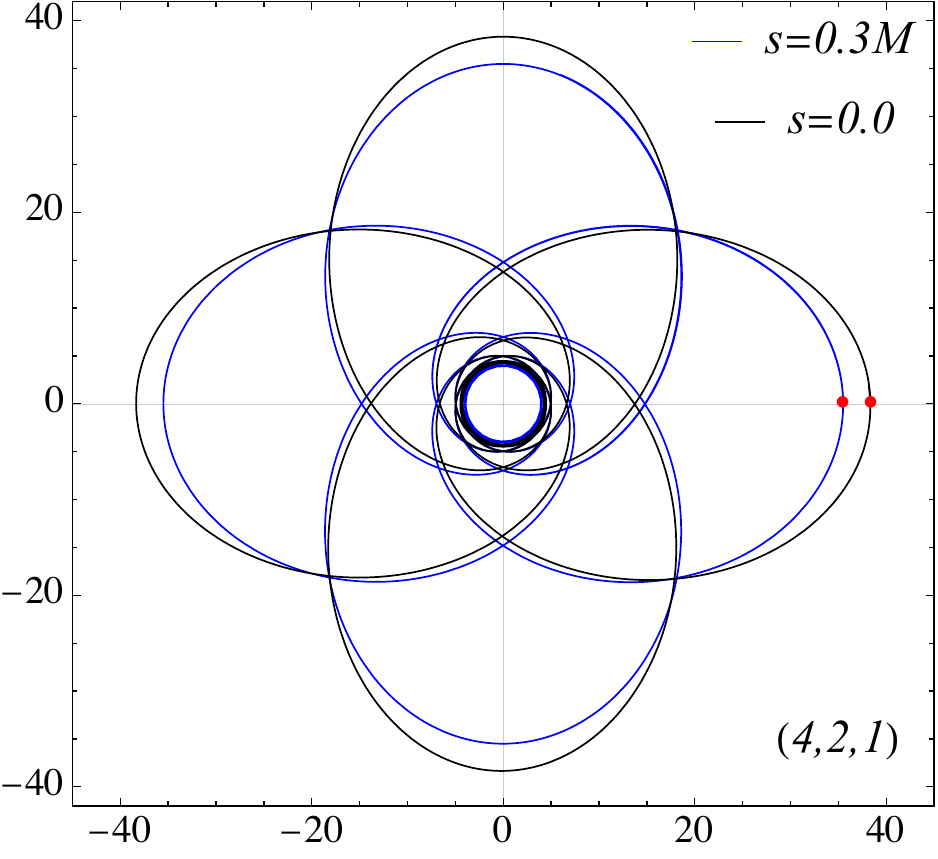}
\caption{Orbits with fixed eccentricity for different taxonomy schemes based on the values given in table (\ref{Table 1}). \label{fig.orbits}}
\end{figure*}

  Also, we show visually the periodic orbits corresponding to the values of the specific energy $\mathcal{E}$ and specific angular momentum $l$ given in Table (\ref{Table 1}) in Fig.(\ref{fig.orbits}) (here we should note to plot periodic orbits we need exact values of the $\mathcal{E}_p$ and $l_p$ given in Appendix (\ref{apdx2})).

{\color{black}\section{Epicyclic frequencies of spinning particles in Schwarzschild spacetime}
\label{sec3}

We now study the fundamental frequencies of nearly circular motion for the aligned-spin sector of the spinning-particle dynamics. In contrast with the geodesic case, the spin--curvature force contributes explicitly to the radial dynamics already at linear order in $s$. Hence the epicyclic frequencies cannot be written only in terms of derivatives of the background metric.


Substituting Eq. \eqref{eq.circular_EL_corrected} into the angular equations of motion, the azimuthal frequency measured by a distant observer becomes
\begin{equation}
\Omega_{\phi}
\equiv
\left.\frac{d\phi}{dt}\right|_{r=r_0}
=
\sqrt{\frac{M}{r_0^3}}
-\frac{3Ms}{2r_0^3}
+\mathcal{O}(s^2).
\label{eq:Omega_phi_corrected}
\end{equation}
This reduces to the Keplerian Schwarzschild frequency in the geodesic limit $s\to0$.  Hereafter, we should note we take whole expressions for circular orbits energy and circular orbits angular momentum before taking approximation up to first order $s$ to avoid confusion.

To determine the radial epicyclic frequency, we perturb the circular orbit as
\begin{equation}
r=r_0+\delta r,\qquad |\delta r|\ll r_0 .
\end{equation}
Expanding $\dot r^{\,2}=\mathcal{R}(r)$ around $r_0$ and using Eq. \eqref{eq.circular_EL_corrected}, the leading term is quadratic in $\delta r$. Differentiating with respect to the affine parameter gives
\begin{equation}
\frac{d^2(\delta r)}{d\lambda^2}
+\omega_r^2\,\delta r=0,
\qquad
\omega_r^2=-\frac{1}{2}\mathcal{R}''(r_0).
\label{eq:omega_r_def}
\end{equation}
Using Eq. \eqref{eq.circular_EL_corrected}, the proper-time radial epicyclic frequency becomes
\begin{equation}
\omega_r^2
=
\frac{M(r_0-6M)}
{r_0^3(r_0-3M)}
+
\frac{3sM^{3/2}(r_0-2M)}
{r_0^{7/2}(r_0-3M)^2}
+\mathcal{O}(s^2).
\label{eq:omega_r_corrected}
\end{equation}

The observable radial epicyclic frequency is defined with respect to the coordinate time $t$. Converting from $\lambda$ to $t$ using $\dot t$ evaluated on the circular orbit, we obtain
\begin{equation}
\Omega_r^2
=
\frac{M(r_0-6M)}
{r_0^4}
+
\frac{3sM^{3/2}(r_0+2M)}{r_0^{11/2}}
+\mathcal{O}(s^2).
\label{eq:Omega_r_corrected}
\end{equation}

In the nonspinning limit this reduces to the standard Schwarzschild result
\begin{equation}
\Omega_r^2
=
\Omega_\phi^2
\left(1-\frac{6M}{r_0}\right).
\end{equation}

Because the Schwarzschild spacetime is spherically symmetric, a small rigid tilt of an aligned-spin circular orbit is equivalent to a rotation of the orbital plane. Therefore, within the strictly aligned-spin sector considered here, the vertical epicyclic frequency coincides with the azimuthal frequency,
\begin{equation}
\Omega_\theta=\Omega_\phi .
\label{eq:Omega_theta_corrected}
\end{equation}
For generic spin orientations this equality need not hold without including spin-precession effects; here it is a consequence of the aligned-spin restriction.

The ISCO is determined by the marginal-stability condition $\Omega_r^2=0$. Solving Eq. \eqref{eq:Omega_r_corrected} perturbatively gives
\begin{equation}
r_{\rm ISCO}
=
6M
-
2\sqrt{\frac{2}{3}}\,s
+\mathcal{O}(s^2),
\label{eq:ISCO_corrected}
\end{equation}
which paves way to find ISCO energy $\mathcal{E}_{ISCO}$ and ISCO angular momentum $l_{ISCO}$ using Eqs. \eqref{eq.circular_EL_corrected} as:
\begin{subequations}
    \begin{align}
        &\mathcal{E}_{ISCO}=\frac{2\sqrt{2}}{3}-\frac{s}{36\sqrt{3}M}\,\\
        &l_{ISCO}=2\sqrt{3}M+\frac{\sqrt{2}s}{3}\,.
    \end{align}
\end{subequations}
Hence positive aligned spin shifts the ISCO inward relative to the Schwarzschild geodesic value $r_{\rm ISCO}=6M$, consistently with the behavior of the effective potential.

Finally, it is useful to distinguish the orbital epicyclic frequencies from the spin-precession frequency. For generic spin orientation, parallel transport of the spin vector introduces an additional precession frequency. This frequency can modulate the signal if the particle spin is not exactly aligned with the orbital angular momentum. In the present aligned-spin analysis, however, the principal observable frequencies are $\Omega_\phi$, $\Omega_r$, and $\Omega_\theta$, with the linear frequencies given by
\begin{equation}
\nu_i=\frac{\Omega_i}{2\pi},
\qquad
i=\phi,r,\theta .
\end{equation}}

\section{Application to quasi-periodic oscillations} \label{sec4}
The epicyclic analysis developed above can be applied directly to quasi-periodic oscillation phenomenology once the coordinate-time frequencies of the spinning particle are known \cite{Witzany:2023bmq}. In the present Schwarzschild problem, the relevant observable frequencies are the orbital frequency $\nu_{\phi}=\Omega_{\phi}/(2\pi)$, the radial epicyclic frequency $\nu_{r}=\Omega_{r}/(2\pi)$, and the vertical epicyclic frequency $\nu_{\theta}=\Omega_{\theta}/(2\pi)$. Since the background is spherically symmetric, the vertical and azimuthal frequencies coincide for the aligned-spin sector considered here, so that $\nu_{\theta}=\nu_{\phi}$ at linear order in the particle spin.

This identification allows us to connect the spinning-particle dynamics to standard QPO constructions. In the relativistic precession model, one identifies the upper and lower high-frequency QPOs with \cite{Stella:1999prl}
\begin{equation}
\nu_{\rm U}=\nu_{\phi},\qquad \nu_{\rm L}=\nu_{\phi}-\nu_{r},
\label{4.1}
\end{equation}
while the nodal precession frequency is
\begin{equation}
\nu_{\rm nod}=\nu_{\phi}-\nu_{\theta},
\label{4.2}
\end{equation}
which is the standard precession-model quantity \cite{Stella:1998lense}. For the non-rotating black hole considered here, Eq. \eqref{4.2} vanishes because of spherical symmetry, and the only nontrivial precessional splitting is the periastron precession frequency
\begin{equation}
\nu_{\rm per}=\nu_{\phi}-\nu_{r}.
\label{4.3}
\end{equation}
Therefore, the principal effect of the particle spin in this setting is to shift the orbital and radial frequencies and, through them, the separation of the two QPO peaks.

The same frequency set can also be used in resonance-based models \cite{Kluzniak:2001ar}. In that case one imposes a commensurability condition of the form
\begin{equation}
p\,\nu_{r}=q\,\nu_{\theta},
\label{4.4}
\end{equation}
with coprime integers $p$ and $q$. Because $\nu_{\theta}=\nu_{\phi}$ in the present case, the resonance condition reduces to
\begin{equation}
p\,\nu_{r}=q\,\nu_{\phi}.
\label{4.5}
\end{equation}
The commonly discussed $3:2$ resonance \cite{Abramowicz:2001bi} is then obtained by setting either $3\nu_{r}=2\nu_{\phi}$ or $2\nu_{r}=3\nu_{\phi}$, depending on the chosen convention for the upper and lower peaks. The resulting resonant radius is shifted by the spin-curvature coupling, since both $\nu_{\phi}$ and $\nu_{r}$ receive linear corrections in $s$.

Operationally, the frequency map is obtained by first solving the circularity conditions for $\mathcal{E}(r_0,s)$ and $l(r_0,s)$ and then substituting these quantities into $\Omega_\phi$ and $\Omega_r$. One then substitutes them into the coordinate-time frequencies $\Omega_{\phi}$ and $\Omega_{r}$. The observable QPO frequencies follow from division by $2\pi$, and the chosen QPO model then provides the required combinations, such as Eqs. \eqref{4.1} and \eqref{4.3} for the relativistic precession model or Eq. \eqref{4.5} for the resonance model.

It is important to emphasize that this construction is kinematical. The Witzany framework supplies the exact spinning-particle frequencies, but a full comparison with X-ray data still requires an additional prescription for how the oscillating trajectory modulates the emitted radiation. Nevertheless, the formalism is sufficient to determine how the spin of the orbiting compact probe shifts the frequency map, the resonant radii, and the inferred source parameters within particle-based QPO models \cite{Witzany:2023bmq}.

The extension is not limited to strictly circular motion. Since the Witzany formalism provides the full bound dynamics in terms of quadratures, and in Schwarzschild space-time even in terms of elliptic functions, one may also define averaged coordinate-time frequencies for eccentric spinning orbits. This opens the possibility of studying QPO scenarios in which the observed peaks are associated not only with local epicyclic oscillations near a circular orbit, but also with mildly eccentric bound motion and its precessional structure.

\section{Lyapunov instability of unstable circular orbits and implications
for transient QPOs}
\label{sec5}

The epicyclic analysis of Sec. \ref{sec3} identifies stable
circular orbits through $\Omega_r^2>0$. The complementary branch, for which
$\Omega_r^2<0$, corresponds to radially unstable circular orbits. In the
geodesic Schwarzschild limit this branch occupies the interval
$3M<r_0<6M$, where the lower endpoint is approached only in the null
photon-sphere limit and the upper endpoint is the ISCO. For spinning
particles the ISCO and the unstable branch are deformed by the
spin--curvature coupling within the leading-in-spin MPD dynamics
\cite{Witzany:2023bmq}.

Unstable circular orbits are important because they control the local
separatrix dynamics of bound orbits. In particular, nearly homoclinic
zoom--whirl trajectories can spend many azimuthal cycles close to an
unstable circular orbit before peeling away. The rate at which nearby
trajectories diverge is governed by the Lyapunov exponent, which also
plays a central role in the geodesic interpretation of black-hole
quasinormal modes in the eikonal limit
\cite{Cardoso:2008bp}. Near-threshold zoom--whirl behavior is likewise
controlled by unstable circular orbits in eccentric black-hole merger
dynamics \cite{Pretorius:2007jn}. This provides a useful connection
between spinning-particle dynamics, transient orbital coherence, and QPO
phenomenology.

Let $r_0$ denote an unstable circular orbit satisfying
\begin{equation}
\mathcal{R}(r_0)=0,
\qquad
\mathcal{R}'(r_0)=0,
\qquad
\mathcal{R}''(r_0)>0,
\end{equation}
where $\mathcal{R}(r)$ is the radial function defined in
Eq. \eqref{eq.R}. For a small perturbation
$r=r_0+\delta r$, the radial equation $\dot r^{\,2}=\mathcal{R}(r)$ gives
\begin{equation}
\frac{d^2(\delta r)}{d\lambda^2}
=
\frac{1}{2}\mathcal{R}''(r_0)\,\delta r .
\end{equation}
Thus $\delta r\propto \exp(\pm\lambda_\tau \lambda)$, with the proper-time
Lyapunov exponent
\begin{equation}
\lambda_\tau^2
=
\frac{1}{2}\mathcal{R}''(r_0)
=
-\omega_r^2(r_0,s).
\end{equation}
Using Eq. \eqref{eq:omega_r_corrected}, one obtains
\begin{equation}
\lambda_\tau^2
=
\frac{M(6M-r_0)}
{r_0^3(r_0-3M)}
-
\frac{3sM^{3/2}(r_0-2M)}
{r_0^{7/2}(r_0-3M)^2}
+\mathcal{O}(s^2).
\label{eq:lambda_tau_corrected}
\end{equation}

For an observer at infinity, the relevant instability rate is measured with
respect to the coordinate time $t$. Therefore we define
\begin{equation}
\Lambda(r_0,s)
\equiv
\sqrt{-\Omega_r^2(r_0,s)},
\qquad
\Omega_r^2<0 .
\end{equation}
This definition follows the standard interpretation of the coordinate-time
Lyapunov exponent as the inverse instability timescale measured by an
asymptotic observer \cite{Cardoso:2008bp}.
Substituting Eq. \eqref{eq:Omega_r_corrected}, we find
\begin{equation}
\Lambda^2
=
\frac{M(6M-r_0)}{r_0^4}
-
\frac{3sM^{3/2}(r_0+2M)}
{r_0^{11/2}}
+\mathcal{O}(s^2).
\label{eq:Lyapunov_corrected}
\end{equation}
Equivalently, away from the ISCO where the square-root expansion is
regular,
\begin{equation}
\Lambda(r_0,s)
=
\Lambda_0(r_0)
\left[
1
-
\frac{3sM^{1/2}(r_0+2M)}{2r^{3/2}(6M-r_0)}
\right]
+\mathcal{O}(s^2),
\label{eq:Lyapunov_expanded_corrected}
\end{equation}
where
\begin{equation}
\Lambda_0(r_0)
=
\frac{\sqrt{M(6M-r_0)}}{r_0^2}
\end{equation}
is the geodesic Schwarzschild result.

Equation \eqref{eq:Lyapunov_corrected} shows that, for positive aligned
spin, the linear spin correction to $\Lambda^2$ is negative throughout the
physical unstable branch. Thus aligned spin weakens the radial instability
at fixed $r_0$. This is the Lyapunov counterpart of the inward ISCO shift:
solving $\Lambda^2=0$ perturbatively gives
\begin{equation}
r_{\rm ISCO}
=
6M
-
2\sqrt{\frac{2}{3}}\,s
+\mathcal{O}(s^2),
\end{equation}
in agreement with Eq. \eqref{eq:ISCO_corrected}.

It is useful to introduce dimensionless variables
\begin{equation}
x=\frac{r_0}{M},
\qquad
\sigma=\frac{s}{M}.
\end{equation}
Then Eq. \eqref{eq:Lyapunov_corrected} becomes
\begin{equation}
M^2\Lambda^2
=
\frac{6-x}{x^4}
-
\sigma
\frac{3(x+2)}{x^{11/2}}
+\mathcal{O}(\sigma^2).
\label{eq:dimensionless_Lambda}
\end{equation}

\subsection{Instability exponent per azimuthal radian}

For zoom--whirl dynamics it is convenient to compare the radial instability
rate with the azimuthal frequency. This ratio measures the instability per
azimuthal radian and is directly related to the number of whirl cycles
accumulated near the separatrix \cite{Pretorius:2007jn,Levin:2008mq}.
We therefore define
\begin{equation}
\gamma(r_0,s)
\equiv
\frac{\Lambda(r_0,s)}{\Omega_\phi(r_0,s)} .
\end{equation}
Using Eqs. \eqref{eq:Omega_phi_corrected} and
\eqref{eq:Lyapunov_corrected}, one obtains
\begin{equation}
\gamma(x,\sigma)
=
\gamma_0(x)
\left[
1+\sigma\,\Pi(x)
\right]
+\mathcal{O}(\sigma^2),
\label{eq:gamma_corrected}
\end{equation}
where
\begin{equation}
\gamma_0(x)
=
\sqrt{\frac{6}{x}-1}
\end{equation}
and
\begin{equation}
\Pi(x)
=\frac{3}{2x^{3/2}}\left(1-\frac{2+x}{6-x}\right).
\label{eq:Pi_corrected}
\end{equation}
For $2<x<6$, the function $\Pi(x)$ is negative. Hence positive aligned
spin increases $\gamma$ at fixed radius, meaning that the radial
perturbation grows more faster per azimuthal radian.

Two limiting cases are useful. First, in the geodesic limit near the
event horizon , $x\to2^+$, one finds $\gamma_0\to\sqrt{2}$.  Also, near the ISCO the
expanded expression for $\Lambda$ in Eq. \eqref{eq:Lyapunov_expanded_corrected}
is singular, while $\Lambda^2$ in Eq. \eqref{eq:Lyapunov_corrected} remains
the correct quantity to use. For plotting, one should therefore plot
$\Lambda=\sqrt{\Lambda^2}$ using Eq. \eqref{eq:Lyapunov_corrected}, with
the domain truncated at the spin-shifted ISCO.

\subsection{Zoom--whirl separatrix dynamics}

In the bound-orbit taxonomy of Levin and Perez-Giz \cite{Levin:2008mq},
orbits with large whirl number accumulate near the separatrix. The same
near-separatrix mechanism underlies the logarithmic scaling of the whirl
phase in near-threshold black-hole encounters \cite{Pretorius:2007jn}.
The separatrix asymptotes to an unstable circular orbit, and the residence
time near this orbit scales as
\begin{equation}
T_{\rm res}
\sim
\frac{1}{\Lambda}\ln\frac{1}{\delta},
\end{equation}
where $\delta$ measures the dimensionless phase-space distance from the
separatrix. During this residence time, the number of accumulated azimuthal
cycles is approximately
\begin{equation}
N_{\rm whirl}(r_0,s)
\simeq
\frac{1}{2\pi\gamma(r_0,s)}
\ln\frac{1}{\delta}.
\label{eq:whirl_corrected}
\end{equation}
Because positive aligned spin increases $\gamma$, it decreases
$N_{\rm whirl}$ at fixed separatrix proximity $\delta$. Thus the
spin--curvature coupling mildly shortens the whirl phase of near-separatrix
trajectories.

\subsection{Implications for transient QPOs}

The Lyapunov exponent also controls the damping envelope of transient
features generated by matter orbiting close to the unstable branch. Since
the same coordinate-time frequency map enters relativistic-precession and
epicyclic-resonance models of high-frequency QPOs
\cite{Stella:1998lense,Stella:1999prl,Kluzniak:2001ar,Abramowicz:2001bi},
the instability rate provides a useful estimate of the coherence time of
near-separatrix transient features. The
coordinate-time instability timescale is
\begin{equation}
T_\Lambda(r_0,s)
=
\frac{1}{\Lambda(r_0,s)} ,
\end{equation}
while the orbital period is
\begin{equation}
T_\phi(r_0,s)
=
\frac{2\pi}{\Omega_\phi(r_0,s)} .
\end{equation}
Therefore the approximate number of coherent orbital cycles before radial
dephasing is
\begin{equation}
N_{\rm coh}
\sim
\frac{T_\Lambda}{T_\phi}
=
\frac{1}{2\pi\gamma}.
\end{equation}
Since aligned spin increases $\gamma$, it decreases the coherence of
transient zoom--whirl features.

On the stable side of the frequency map, resonance-based QPO models often
use commensurability conditions of the form \cite{Kluzniak:2001ar,Abramowicz:2001bi}
\begin{equation}
p\,\Omega_r-q\,\Omega_\phi=0 .
\end{equation}
Near the separatrix, the continuation of the radial mode into the unstable
branch replaces the real radial frequency by the Lyapunov rate
$\Lambda=\sqrt{-\Omega_r^2}$. A simple phenomenological estimate for the
radial width of a transient feature is therefore
\begin{equation}
\Delta r
\sim
\frac{\Lambda}
{\left|\partial_r(p\Omega_r-q\Omega_\phi)\right|_{r=r_{\rm res}}},
\end{equation}
where the derivative is evaluated on the stable side of the resonance.
Thus the spin dependence of $\Lambda$ affects not only the central
frequency map but also the expected width and coherence of transient
QPO-like structures.

\begin{figure*}[t]
\includegraphics[width=0.46\textwidth]{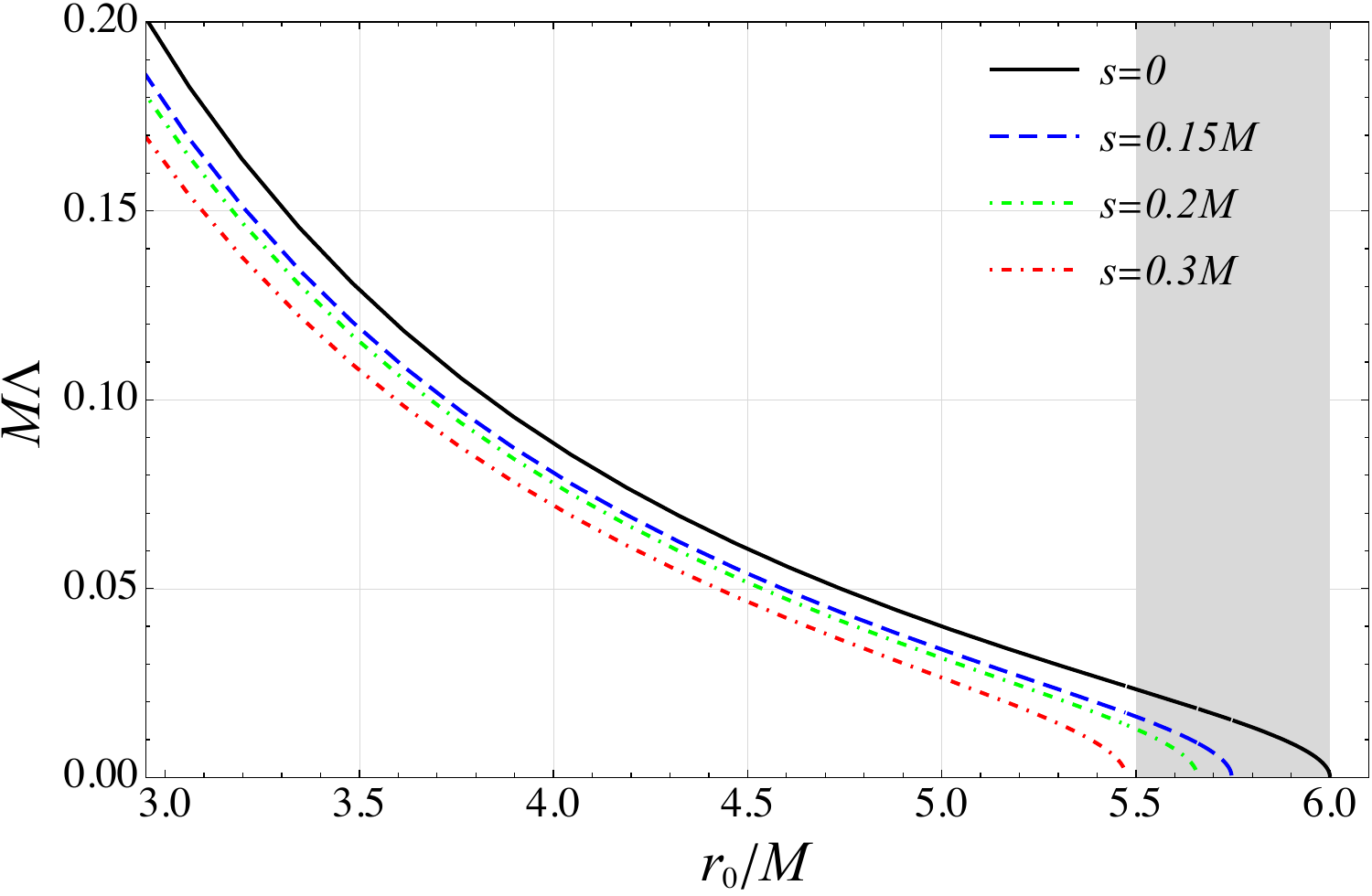}
\hspace{0.5cm}
\includegraphics[width=0.46\textwidth]{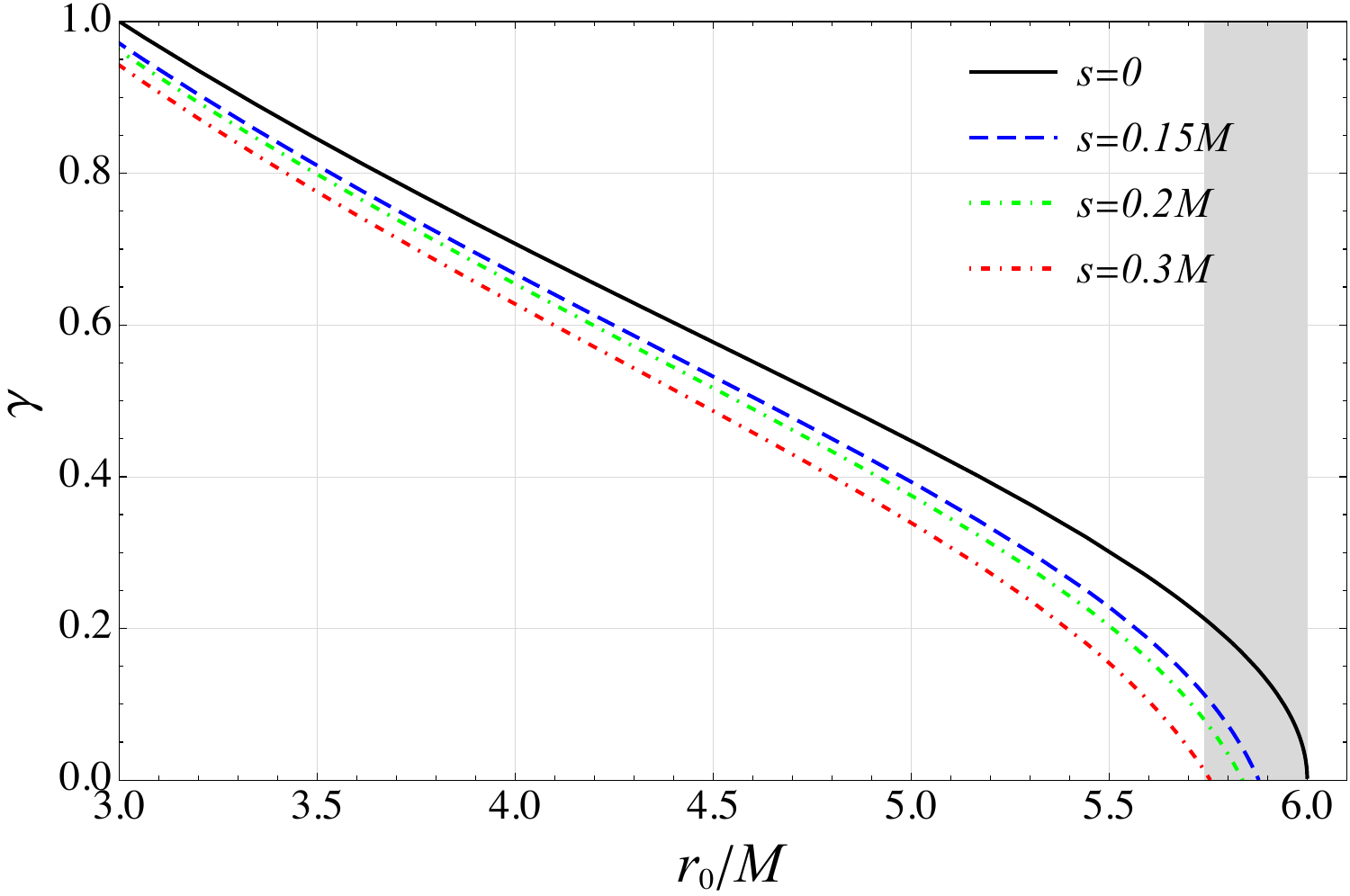}
\caption{\label{Fig.Lyap}
Dimensionless Lyapunov exponent $M\Lambda$ (left panel) and instability
exponent per azimuthal radian $\gamma=\Lambda/\Omega_\phi$ (right panel)
for unstable circular orbits of spinning particles in Schwarzschild
spacetime, computed from the corrected expressions
Eqs. \eqref{eq:Lyapunov_corrected} and \eqref{eq:Omega_phi_corrected}.
Curves are shown as functions of $r_0/M$ for representative aligned spins
$s/M\in\{0,\,0.15,\,0.20,\,0.30\}$.  The right endpoint of each curve is
the spin-shifted ISCO at which $\Lambda$ vanishes; the shaded band marks
the migration of the marginal orbit from $r_{\rm ISCO}(0)=6M$ to the
numerically determined $r_{\rm ISCO}(0.30M)\simeq 5.5101M$. }
\end{figure*}

\begin{table}[h]
\centering
\begin{tabular}{cccc}
\hline\hline
$s/M$ & $r_{\rm ISCO}/M$ (numerical) & $r_{\rm ISCO}/M$ (linear) & difference \\
\hline
$0.00$ & $6.000$ & $6.000$ & $\phantom{-}0.000$ \\
$0.10$ & $5.829$ & $5.837$ & $-0.008$ \\
$0.15$ & $5.736$ & $5.755$ & $-0.019$ \\
$0.30$ & $5.419$ & $5.510$ & $-0.091$ \\
$0.45$ & $4.990$ & $5.265$ & $-0.275$ \\
\hline\hline
\end{tabular}
\caption{\label{tab:ISCO_comparison}
Comparison of the linear-in-$s$ ISCO estimate of
Eq. \eqref{eq:ISCO_corrected} with the outer numerical root of the corrected
condition $\Lambda^{2}(r_0,s)=0$ from Eq. \eqref{eq:Lyapunov_corrected}.
The linear estimate is accurate to better than $0.02M$ for $s/M\lesssim
0.15$, consistent with the regime of validity of the pole--dipole
expansion, and progressively overestimates $r_{\rm ISCO}$ at larger
spin.}
\end{table}

Figure \ref{Fig.Lyap} displays the dimensionless quantities $M\Lambda$
and $\gamma$ for representative aligned spins.  The $s=0$ curves reproduce
the geodesic Schwarzschild result, with $\gamma_0(x)=\sqrt{6/x-1}$
approaching unity in the photon-sphere limit and vanishing at the ISCO.
Turning on positive aligned spin suppresses the instability throughout
the perturbatively reliable part of the unstable branch.   

\section{Unbound orbits}\label{sec6}
\begin{figure*}[t]
\includegraphics[width=0.33\textwidth]{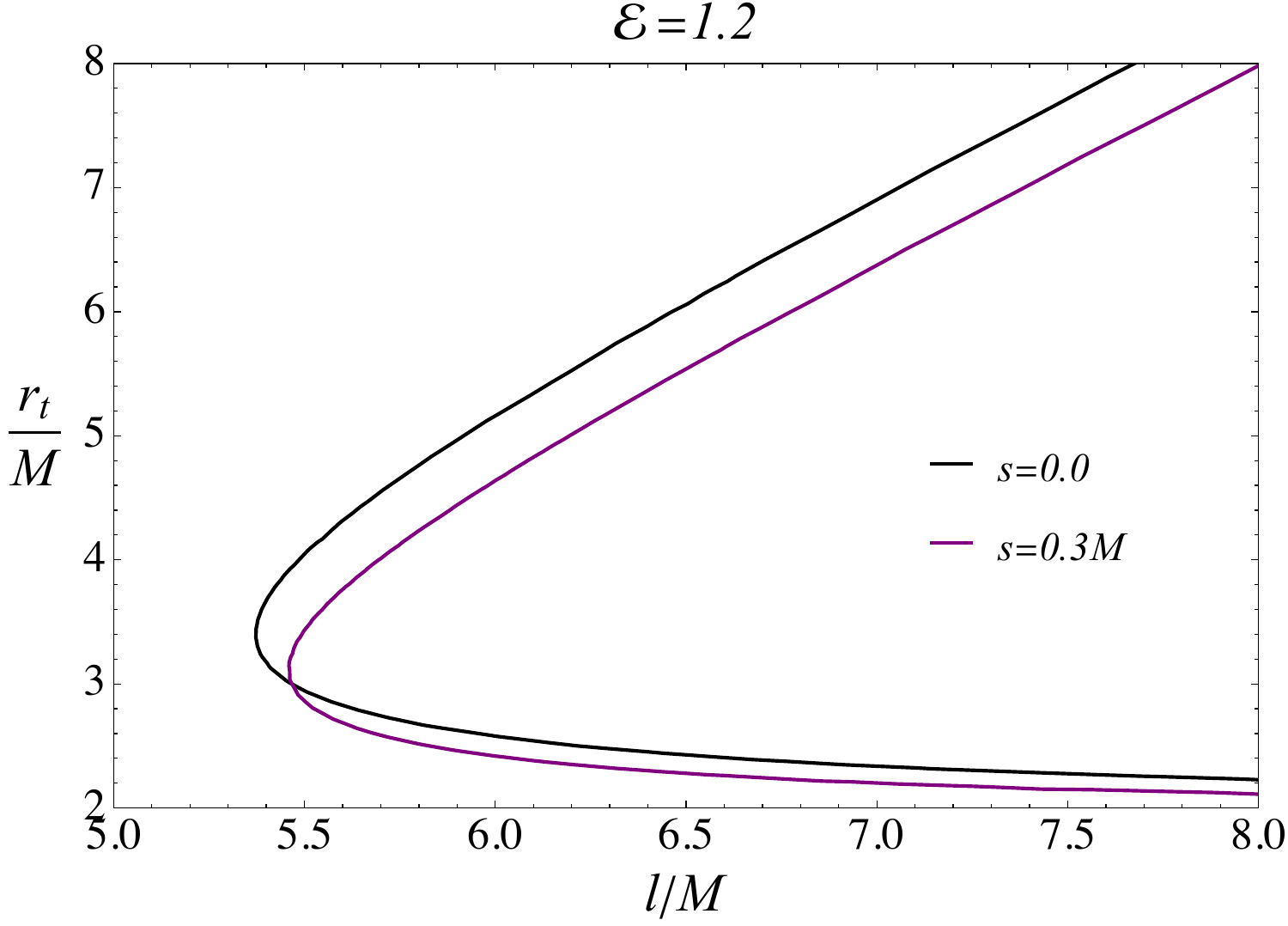}
\includegraphics[width=0.325\textwidth]{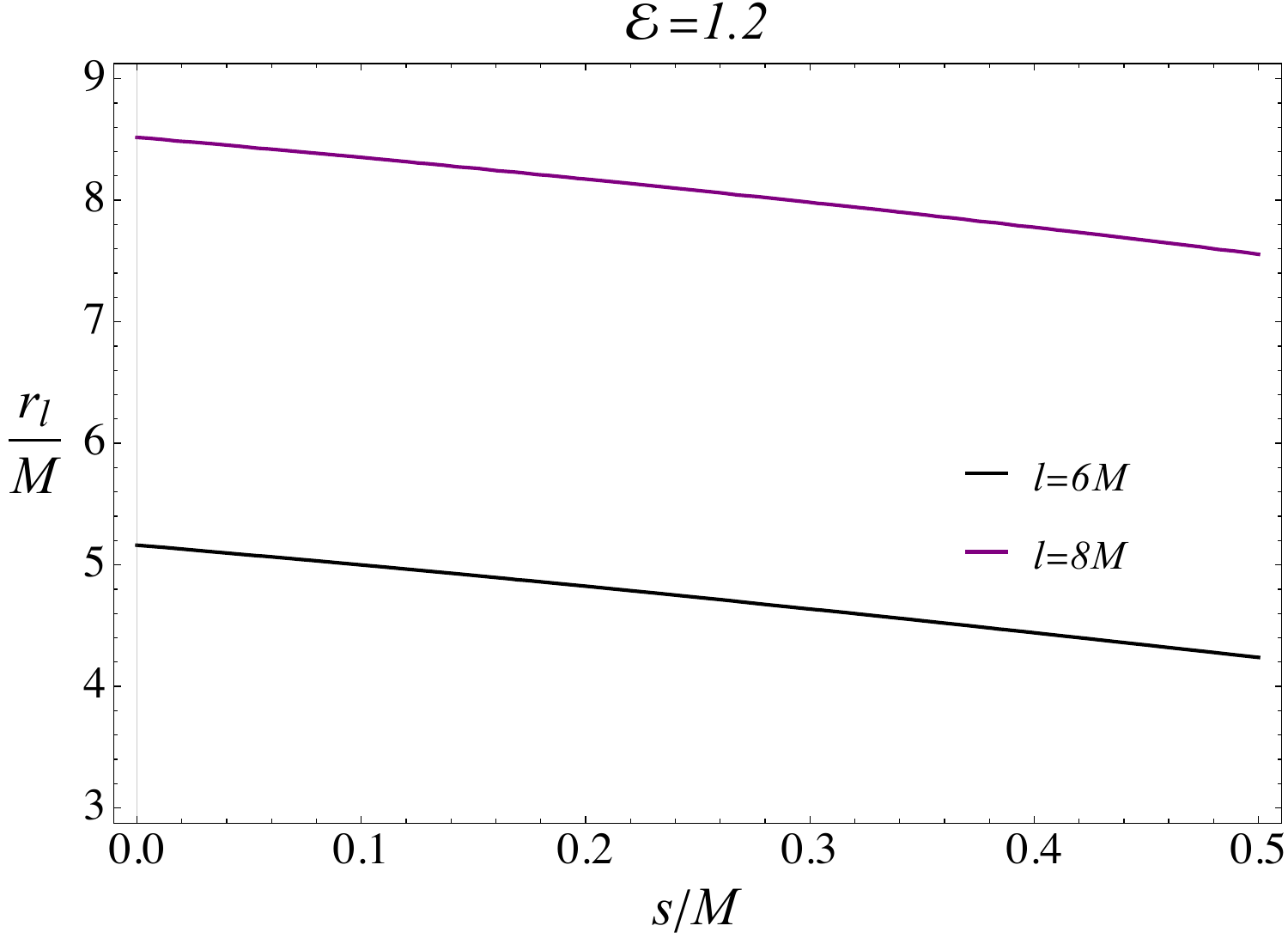}
\includegraphics[width=0.33\textwidth]{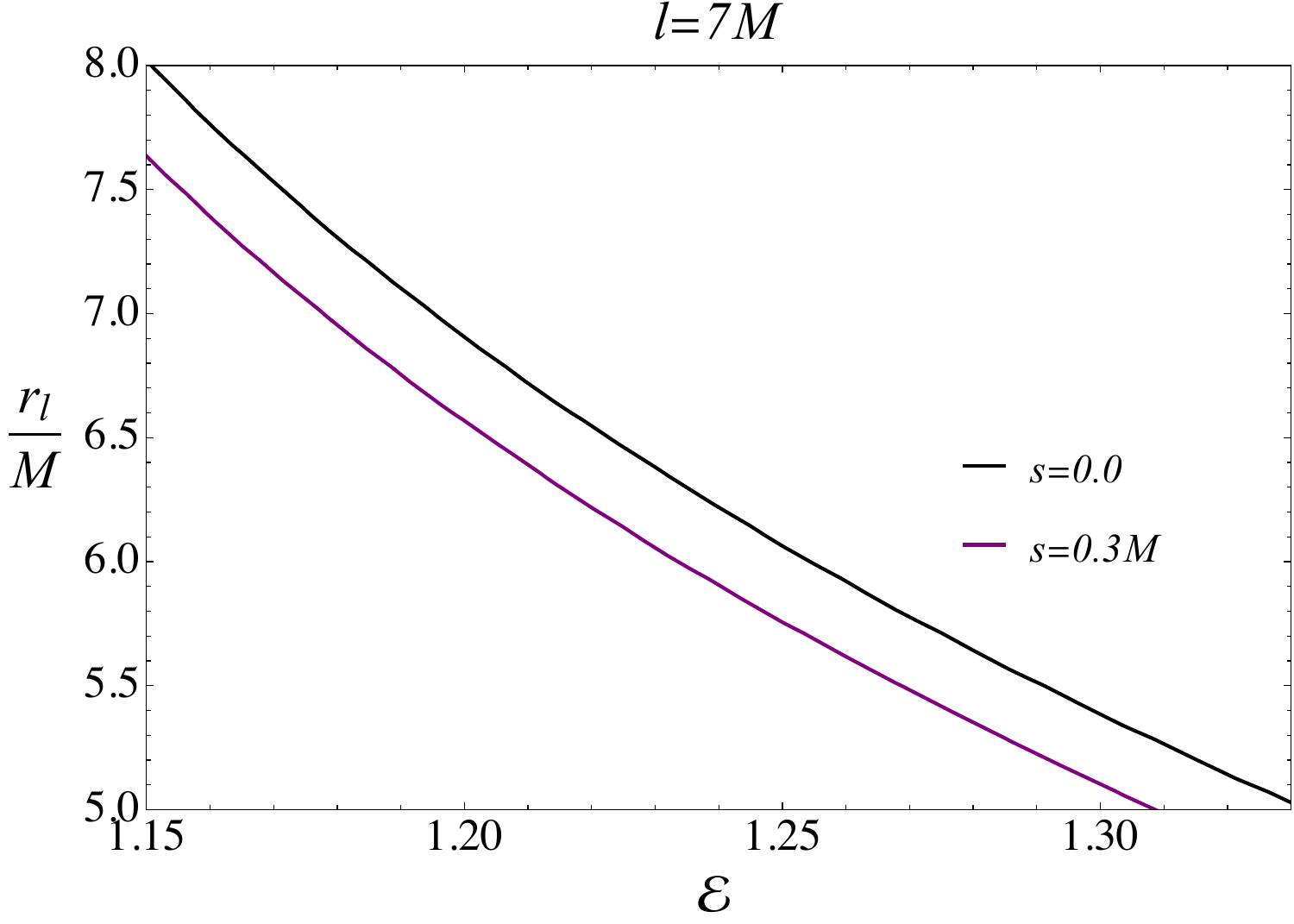}
\caption{
 The relationship between the turning point radii $r_t$ and specific angular momentum at constant specific energy $\mathcal{E}$ is shown in the left panel; the middle panel illustrates how the maximum orbital separation $r_l$ varies with spin $s$; and the right panel depicts the dependence of $r_l$ on specific energy $\mathcal{E}$. \label{fig.rt}}
\end{figure*}

From the asymptotic form of the radial equation, the necessary condition for an unbound orbit is $\mathcal{E}>1 .$ The distinction between scattering and capture is not determined by this inequality alone, but by the root structure of the radial polynomial $P(r)$. In particular, scattering orbits possess an outer positive turning point, whereas capture trajectories cross the horizon after reaching the inner branch of the radial potential. Also,  the turning point $r_t$ of the spinning particle can be found from the condition $\frac{dr}{d\phi}=0$ in Eq.(\ref{eq.dr}). Then, we illustrate dependence of the turning point on the specific energy $l$ of the spinning particle in Fig.(\ref{fig.rt}). For the test particle, two turning points exist —  the smallest orbital separation $r_s$ and the largest orbital separation $r_l$ (the left panel of Fig.(\ref{fig.rt})). In the case of unbound trajectories, designated as orbits of the first kind (OFK), any particle that reaches $r_l$ will then move outward to infinity following a hyperbolic motion. The dependence of the $r_l$ on the particle spin $s$ (middle panel of Fig.(\ref{fig.rt})) reveals that increasing the value of the spin of the particle causes decreasing the value of the $r_l$. If particles reach to the $r_s$, they are captured by the event horizon along a spiral path (orbits of the second kind (OSK)). Below, we provide two kind of orbits separately.
  
\subsection{OFK}
The one of the roots of the Eq.(\ref{eq.dr1}) is negative for OFK, so for the spinning particle starting motion from $r=r_1$ and $\phi=0$ Eq.(\ref{eq.phi2}) can be rewritting as:
\begin{eqnarray}\label{eq.OFK phi}
    \phi(r)=\frac{2\left(l+s\mathcal{E}\right)}{\sqrt{(1-\mathcal{E}^2)(r_1-r_3)r_2}}\left[F(x_0,k_0)-K(k_0)\right]\,,
\end{eqnarray}
which paves a way to obtain the trajectory of the spinning particle as:
\begin{eqnarray}\label{eq.OFK rphi}
    r(\phi)=\frac{r_2(r_1-r_3)-r_3(r_1-r_2)\text{sn}^2\left(\phi_1,k_0\right)}{r_1-r_3-(r_1-r_2)\text{sn}^2\left(\phi_1,k_0\right)}\,,
\end{eqnarray}
in which $\phi_1=\phi_0+K(k_0)$.
\begin{figure*}[t]
\includegraphics[width=0.45\textwidth]{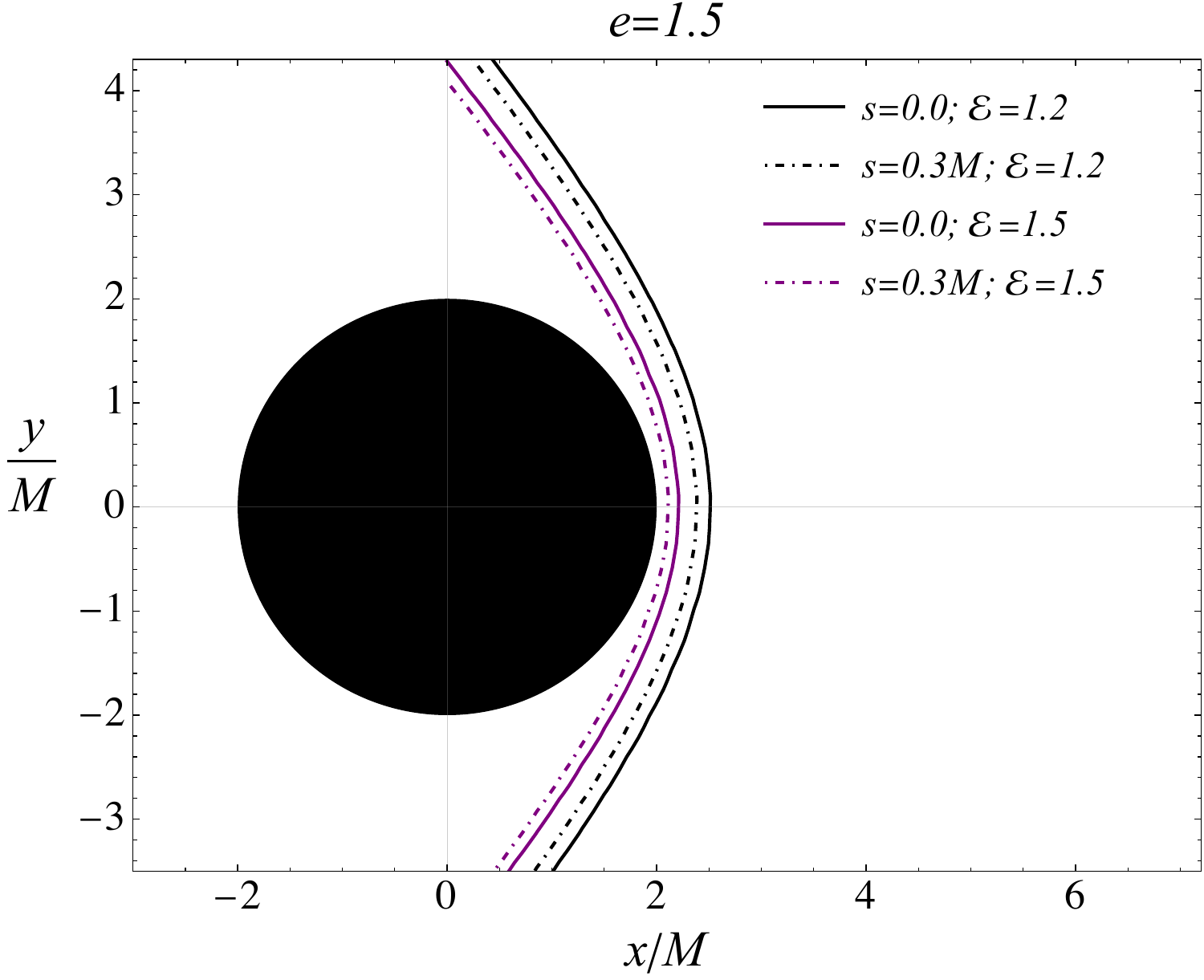}
\includegraphics[width=0.45\textwidth]{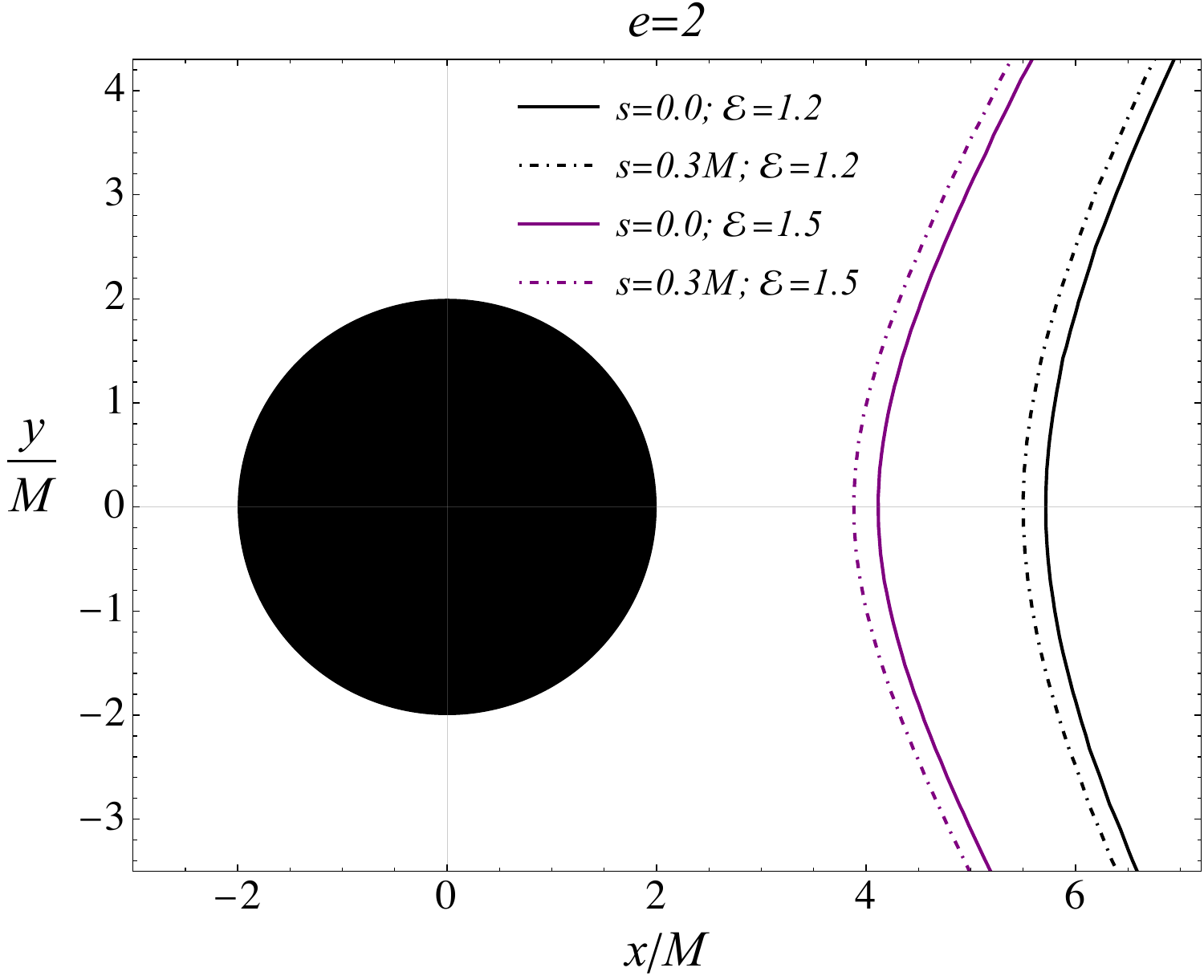}
\caption{
 The gravitational Rutherford scattering plotted for different cases. \label{fig.hyp}}
\end{figure*}

Then we have shown the OFK fo the spinning particle in Fig.(\ref{fig.hyp}). Similarly to previous analysis, it is clear from Fig.(\ref{fig.hyp}) that the presence of the spin of the particle $s$ causes test particles come closer to the black hole.

\subsection{OSK}

A particle that attains $r_s$ is subsequently pulled into the event horizon through a spiraling motion, a process we classify as an OSK in Sec. \ref{sec6}. Again, for spinning particles beginning their motion  at the point $r = r_3$ and $\phi = 0$, Eq. (\ref{eq.phi}) can be integrated to obtain:
\begin{eqnarray}\label{eq.OSK phi}
    \phi(r)=\frac{2(l+s\mathcal{E})}{\sqrt{(1-\mathcal{E}^2)(r_1-r_3)r_2}}F(x_1,k_1)\,,
\end{eqnarray}
but for OSK corresponding modulus and argument are $k_1=\sqrt{\frac{(r_1-r_3)r_2}{(r_1-r_2)r_3}}$ and $x_1=\arcsin{\sqrt{\frac{(r_1-r_2)(r-r_3)}{(r_1-r_3)(r-r_2)}}}$, respectively.
Then we can find analytical solution for the OSK as:
\begin{eqnarray}\label{eq.rphi OSK}
    r(\phi)=\frac{r_3(r_1-r_2)-r_2(r_1-r_3)\text{sn}^2(\phi_0,k_1)}{r_1-r_2-(r_1-r_3)\text{sn}^2(\phi_0,k_1)}\,.
\end{eqnarray}

\begin{figure*}[t]\centering
\includegraphics[width=0.5\textwidth]{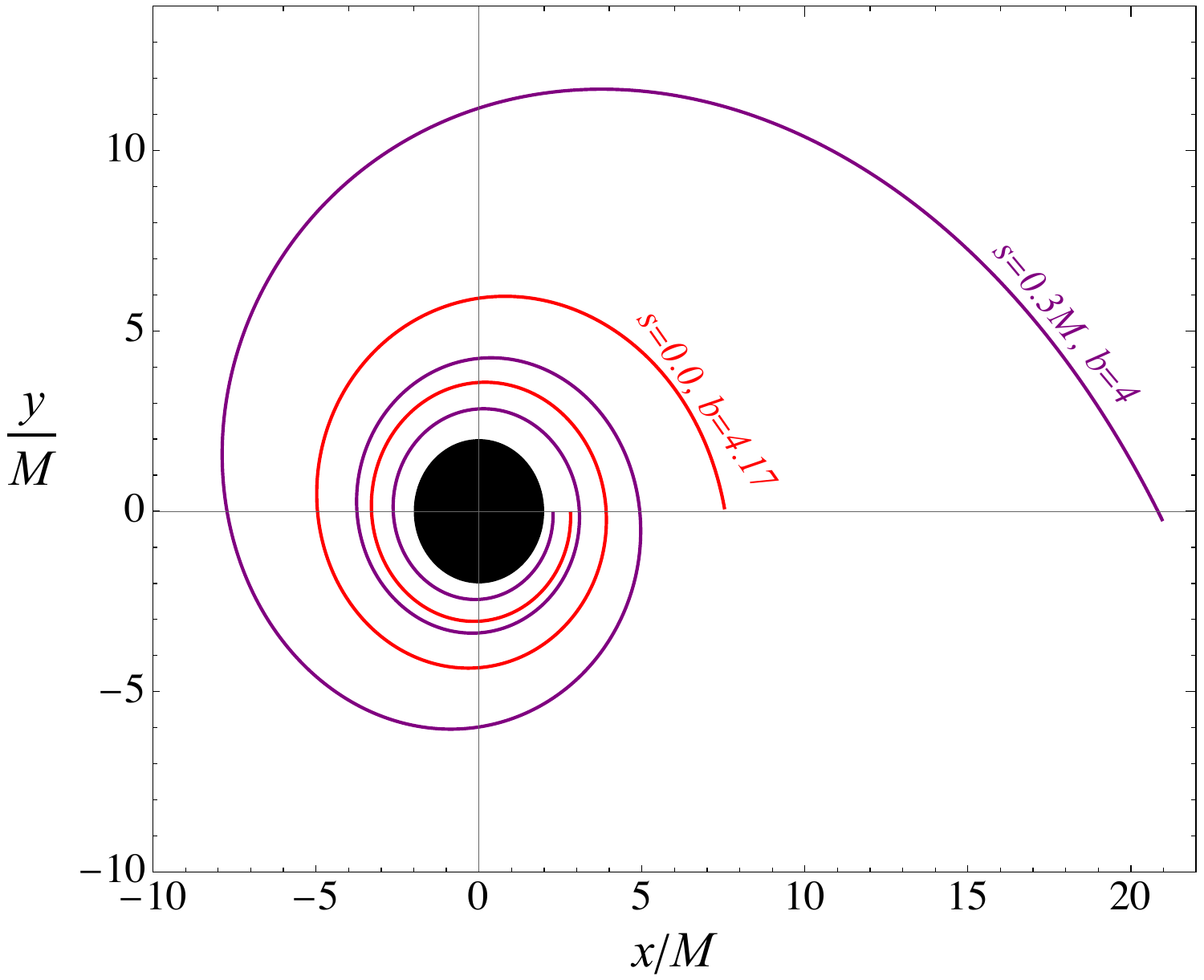}
\caption{OSK as functions of the spin parameter $s$ and the impact parameter $b = l/\mathcal{E}$ \label{fig.OSK}}
\end{figure*}

The the OSK is illustrated for spinning particle in Fig.(\ref{fig.OSK}) for different values of the impact parameter $b=\frac{l}{\mathcal{E}}$.

\subsection{Constraints on the spin $s$ parameter from periastron precession.}

Perihelion precession of the spinning particle during one revolution  in the vicinity of the Schwarzschild BH can be found using condition:
\begin{eqnarray}\label{eq.pres. spinning}
  \delta\phi=2(\phi(r_1)-\pi)\,,  
\end{eqnarray}
where $\phi$ is given in Eq.(\ref{eq.phi2}). Again till first order of the spin $\mathcal{O}(s^2)$ we can write precession as $\delta\phi=\delta\phi^g+s\delta\phi$, where
\begin{subequations}\label{eq.pres.sp 2}
    \begin{align}
        &\delta\phi^g=\frac{4K(k_p)}{\sqrt{1+2\epsilon(e-3)}}-2\pi\,,\\
        &\delta\phi=\frac{2}{\lambda}\sqrt{\frac{(1-2\epsilon)^2-4\epsilon^2e^2}{1+2\epsilon(e-3)}}\left[\frac{E(k_p)}{1-2\epsilon(e+3)}-K(k_0)\right]\,,
    \end{align}
\end{subequations}
in which:
\begin{eqnarray}
    k_p=\frac{4e\epsilon}{1+2\epsilon(e-3)}\,
\end{eqnarray}
and $\epsilon=\frac{M}{\lambda}$. For weak-field regime $\lambda\gg M$ or $\epsilon\ll1$ Eq.(\ref{eq.pres.sp 2}) can be rewritten as:
{\color{black}In the weak-field regime $\lambda\gg M$, the periastron advance takes the schematic expansion
\begin{equation}
\delta\phi
=
\frac{6\pi M}{\lambda}\left[1-\frac{s}{\lambda}\right]
+
\mathcal{O}\!\left(\frac{M^2}{\lambda^2},
\frac{s\sqrt{M}}{\lambda^{3/2}},
s^2\right).
\label{eq:weak_precession_corrected}
\end{equation}
The first term is the standard Schwarzschild geodesic contribution, while the second term is the leading spin--curvature correction for the aligned-spin convention used in the present work.
This expression displays explicitly the leading spin correction to the periastron precession.

We emphasize that Eq. \eqref{eq:weak_precession_corrected} is valid within the pole--dipole approximation and for sufficiently small $s/M$. For large spin values or near the separatrix, the full elliptic-integral expression should be used instead of the weak-field expansion.

To illustrate the numerical method, we take Mercury as a test case with the following parameters:
\begin{subequations}\nonumber
    \begin{align}
        &\frac{GM_\odot}{c^2}=1.47662504\times10^{3} [m]\\\nonumber
        &a=5.7909175\times10^{10}[m]\,,\\\nonumber
        & e=0.20563069\,,
    \end{align}
\end{subequations}
here $a=\frac{\lambda}{(1-e^2)}$ is the semi-major axis of the orbit. Then we can calculate  Mercury's perihelion:
\begin{eqnarray}\label{eq.Mercury prec.}
    \delta\phi&=&2\pi\times(7.98746-\\\nonumber
    &-&2.12665\times10^{-7}\frac{sc^2}{M_\odot G})\times10^{-8}\text{rad/rev}\,,
\end{eqnarray}
also we should note observed value of the Mercury's perihelion as (\cite{Benczik:2002tt}):
\begin{eqnarray}\nonumber
\delta\phi_{obs}=2\pi\times(7.98734\pm0.00037)\times10^{-8}\text{rad/rev}\,.
\end{eqnarray}

Now we find the effect of the spin $s$ of the S2 star on the periastron precession assuming Sgr A* as a Schwarzschild BH and adopting following parameter for our estimation (\cite{GRAVITY:2020gka}):
\begin{subequations}\nonumber
    \begin{align}
        & M_{Sgr A^*}=4.26\times10^6 M_\odot\,\\\nonumber
        &a_{S2}=970 [\text{au}]\,\\\nonumber
        &1 \text{au}=1.4959787\times10^{11} [m]\,\\\nonumber
        & e_{S2}=0.884649\,\\\nonumber
        &T_{S2}=16.052 [\text{years}]\,.
    \end{align}
\end{subequations}
Then we can calculate numerical value of the precession due to the spin $s$ as:
\begin{eqnarray}
    \delta\phi=-\frac{s}{M_{Sgr A^*}}\times0.00962723 \,\,[''/\text{year}]\,,
\end{eqnarray}
and the precession due to pure GR effect:
\begin{eqnarray}
    \delta\phi^g=48.3433\,\,[''/\text{year}]\,,
\end{eqnarray}
so we can write the full precession for S2 star as:
\begin{eqnarray}
    \delta\phi=48.3433-\frac{s}{M_{Sgr A^*}}\times0.00962723 \,\,[''/\text{year}]\,.
\end{eqnarray}
One open question in relativistic astrophysics is the periastron precession of an exoplanetary system. We take WASP-43b as example to calculate perihelion precession (\cite{Bernabo:2025wasp43b}) :
\begin{subequations}
    \begin{align}
        & M_{WASP-43}=0.717M_{\odot}\,,\\\nonumber
        & a=0.01526\,\,[\text{au}]\,,\\\nonumber
        &e=0.00188\,,\\\nonumber
        &T_{WASP-43b}=0.81347\,[\text{days}].
    \end{align}
\end{subequations}
Subsequently, we can calculate WASP-43b's perihelion precession:
\begin{eqnarray}
    \delta\phi_{WASP-43b}&=&22.1582\\\nonumber
    &-&\frac{sc^2}{M_{WASP-43}G}\times0.00010276\,\,[''/\text{days}]\,,
\end{eqnarray}
which is still very small to explain observed value of the WASP-43b's precession $\delta\phi_{obs}=621.72^{+29.88}_{-32.04}\,[''/\text{days}]$ (\cite{Bernabo:2025wasp43b}).

\section{Numerical kludge method for producing gravitational wave signals from periodic orbits.} \label{sec7}

In this section, we analyze the effect of the spin $s$ on the gravitational waveforms from periodic orbits in the vicinity of the BH. We consider EMRI system including a spinning stellar-mass object along the periodic orbit around Schwarzschild BH. The motion generates GWs that could carry information regarding the spin $s$ of the stellar-mass object. The numerical kludge waveform model can be used to investigate the gravitational waveforms produced by the EMRI system. We begin by solving the equations of motion to investigate periodic orbits of the spinning particles around the BH, after which the GWs are generated using the symmetric and trace-free (STF) mass quadrupole formulation of gravitational radiation (\cite{Alloqulov:2026regularBH,Yang:2025qcbh}):

\begin{figure*}[t]
    \centering
    \begin{minipage}{0.47\textwidth}
        \centering
        \includegraphics[width=\textwidth]{orbit_3,0,1.pdf}
        \caption{(3,0,1) periodic orbit}
    \end{minipage}
    \hfill
    \begin{minipage}{0.47\textwidth}
        \centering
        \includegraphics[width=0.8\textwidth]{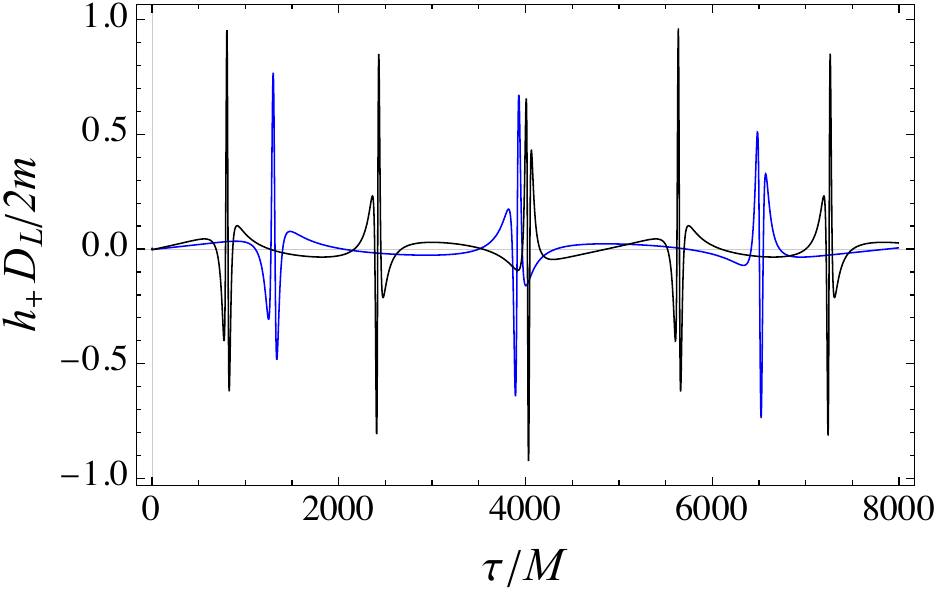}
        
        \vspace{0.3cm}
        
        \includegraphics[width=0.8\textwidth]{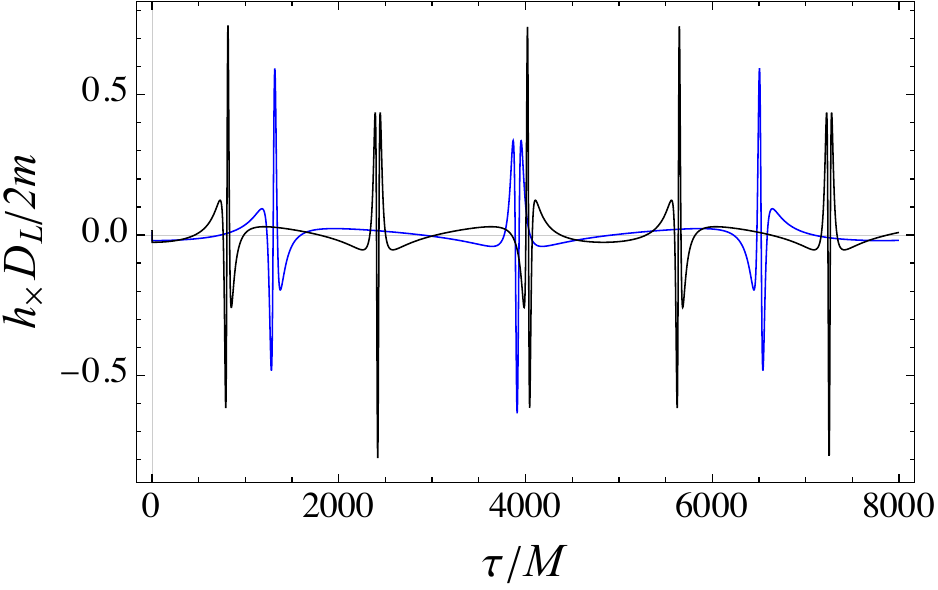}
        \caption{GW waveforms}
    \end{minipage}
    \caption{The (3,0,1) periodic orbit and its resulting GW signal from the EMRI system, featuring a supermassive BH orbited by a small spinning object.}
    \label{fig.GW}
\end{figure*}

\begin{eqnarray}\label{eq.I}
    I^{ij}=\left[\int d^3xx^ix^jT^{tt}(t,x^i)\right]^{(STF)}\,,
\end{eqnarray}
where the $T^{tt}(t,x^i)$ component of the stress-energy tensor for the small celestial body can be expressed as:
\begin{eqnarray}\label{eq.Ttt}
    T^{tt}(t,x^i)=m\delta^3\left(x^i-Z^i(t)\right)\,,
\end{eqnarray}
in which $Z^i(t)$ is the trajectory of the small celestial body.

Subsequently, the metric perturbations can be obtained as:
\begin{eqnarray}\label{eq.h}
    h_{ij}=\frac{2}{D_L}\frac{d^2I_{ij}}{dt^2}=\frac{2m}{D_L}\left(a_ix_j+a_jx_i+2v_iv_j\right)\,,
\end{eqnarray}
here $a_i$ and $v_i$ are the acceleration and velocity of the small object, respectively. Also, the trajectory of the small object is projected into Cartesian coordinate system as:
\begin{eqnarray}\label{eq.transformation}
    x=r\sin{\theta}\cos{\phi}\,,\,\,\,y=r\sin{\theta}\sin{\phi}\,,\,\,\,z=r\cos{\theta}\,,
\end{eqnarray}
 and the luminosity distance between the EMRI system and the detector is given by $D_L$ in Eq.(\ref{eq.h}). In order to match real detector specifications, the calculated GW signal requires projection onto a coordinate system consistent with the detector’s reference framework (\cite{Yang:2025qcbh,PoissonWill:2014}), with the basis taking the following form:
 \begin{subequations}\label{eq.e}
     \begin{align}
         &e_X=\left[\cos{\xi},-\sin{\xi},0\right]\,,\\
         &e_Y=\left[\cos{i}\sin{\xi},\cos{i}\cos{\xi},-\sin{i}\right]\,,\\
         &e_Z=\left[\sin{i}\sin{\xi},\sin{i}\cos{\xi},\cos{i}\right]\,,
     \end{align}
 \end{subequations}
 where $\xi$ represents the longitude of the periastron, and $i$ denotes the inclination angle between the test particle's orbit and the line of sight, while $e_X$, $e_Y$, $e_Z$ form an orthogonal coordinate basis aligned with the detector. Then corresponding GW polarization can be obtained as:
\begin{subequations}\label{eq.h1}
    \begin{align}
        &h_+=\frac{1}{2}\left(e_X^ie_X^j-e_Y^ie_Y^j\right)h_{ij}\,,\\
        &h_\times=\frac{1}{2}\left(e_X^ie_Y^j+e_Y^ie_X^j\right)h_{ij}\,.
    \end{align}
\end{subequations}
 Finally, considering Eq.(\ref{eq.e}) Eq.(\ref{eq.h1}) can be rewritten as:
 \begin{subequations}\label{eq.h2}
     \begin{align}
         &h_+=\frac{1}{2}\left(h_{\xi\xi}-h_{ii}\right)\,,\\
         &h_\times=h_{i\xi}\,,
     \end{align}
 \end{subequations}
 in which:
 \begin{widetext}
     \begin{subequations}
         \begin{align}
             &h_{\xi\xi}=h_{xx}\cos^2{\xi}-h_{xy}\sin{2\xi}+h_{yy}\sin^2{\xi}\,,\\
             &h_{ii}=\cos{i}\left[h_{xx}\sin^2{\xi}+h_{xy}\sin{2\xi}+h_{yy}\cos^2{\xi}\right]+h_{zz}\sin^2{i}-\sin{2i}\left[h_{xz}\sin{\xi}+h_{yz}\cos{\xi}\right]\,,\\
             &h_{i\xi}=\cos{i}\left[\frac{1}{2}h_{xx}\sin{2\xi}+h_{xy}\cos{2\xi}-\frac{1}{2}h_{yy}\sin{2\xi}\right]+\sin{i}\left[h_{yz}\sin{\xi}-h_{xz}\cos{\xi}\right]\,.
         \end{align}
     \end{subequations}
 \end{widetext}

Then, we show the GWs radiated from celestial bodies in periodic orbits around BHs in Fig.(\ref{fig.GW}). We consider an EMRI system characterized by a celestial body mass $m\sim10M_\odot$, an SMBH mass $M\sim10^7M_\odot$, latitude and inclination angle both set to $\xi=i=\frac{\pi}{4}$, and a luminosity distance $D_L=200Mpc$. In Fig.(\ref{fig.GW}), we present the gravitational wave polarizations ($h_+$ and $h_\times$) generated by the (3, 0, 1) periodic orbit. From this figure, we observe both zoom and whirl stages, where the calm oscillations correspond to the zoom phase and the rapid oscillations correspond to the whirl phase. The GW emission exhibits a calm pattern when the small object is on a distant elliptical orbit around the BH, and conversely, a whirl pattern when the object undergoes rapid whirling motion near the BH. One should note that the frequency of the gravitational waves undergoes a dramatic increase in the whirl stage, thereby generating strong oscillations. From Fig. (\ref{fig.GW}), it can be concluded that the presence of the spin $s$ enlarges both the zoom and whirl regions. Consequently, the spin of the stellar-mass object results in a temporal shift of the gravitational waveforms toward larger time values, accompanied by a slight reduction in the values of the polarization components. 
 
\section{Conclusions}
\label{sec8}
We have analyzed spinning test-particle motion around a Schwarzschild black hole within the Mathisson--Papapetrou--Dixon pole--dipole approximation, supplemented by the Tulczyjew--Dixon spin supplementary condition.  By restricting to the equatorial aligned-spin sector and retaining only terms linear in the specific spin $s$, we obtained a controlled analytic description of the radial potential, circular-orbit conditions, ISCO shift, periodic bound trajectories, coordinate-time epicyclic frequencies, and Lyapunov instability of unstable circular orbits.

The spin--curvature interaction changes the circular-orbit energy and angular momentum already at first order in $s$.  In the sign convention used here, positive aligned spin shifts the ISCO inward, with $r_{\rm ISCO}=6M-2\sqrt{2/3}\,s+\mathcal{O}(s^2)$.  This result agrees with the qualitative behavior of the effective potential and shows that the spin of the probe can deform the strong-field orbital structure even when the central black hole is nonrotating.  The statement should be understood within the perturbative pole--dipole regime; curves shown at comparatively large $s/M$ are best regarded as illustrations of trends rather than as controlled approximations.

We also derived the azimuthal and radial epicyclic frequencies measured with respect to Schwarzschild coordinate time.  In the aligned-spin sector considered here, spherical symmetry implies $\Omega_\theta=\Omega_\phi$, whereas the radial epicyclic frequency receives an independent spin-dependent correction.  The resulting frequency map shifts the periastron-precession frequency and the resonance radii entering relativistic-precession and epicyclic-resonance prescriptions for high-frequency QPOs.  These applications are kinematical: a direct comparison with X-ray timing observations would require an additional model for the radiating matter and its modulation mechanism.

For eccentric bound motion, we constructed periodic trajectories and organized them using the Levin--Perez-Giz zoom--whirl taxonomy.  The particle spin changes the energy--angular-momentum values associated with a fixed taxonomy label and fixed eccentricity, thereby deforming the morphology of strong-field periodic orbits.  Since such orbits organize generic bound motion near the separatrix, this construction provides a useful analytic route for studying spin-dependent changes in zoom--whirl behavior and for producing approximate gravitational-wave polarizations from representative trajectories.

Finally, we examined unstable circular orbits through their coordinate-time Lyapunov exponent.  The instability exponent determines the local divergence rate of nearby trajectories and therefore controls the residence time of near-separatrix or near-homoclinic motion.  Within the same linear-in-spin approximation, the Lyapunov analysis links the spin-corrected epicyclic frequency map to the coherence properties of transient zoom--whirl features.  This connection is useful for assessing how finite-size spin effects may enter approximate descriptions of transient strong-field signals.

Several extensions are natural.  Quadratic-in-spin effects and spin-induced quadrupole couplings should be included to go beyond the leading pole--dipole approximation.  Generic spin orientations should also be considered, since spin precession can break the equality $\Omega_\theta=\Omega_\phi$ and introduce additional modulation frequencies.  The corresponding Kerr problem is the next essential step, because frame dragging, black-hole spin, spin--orbit coupling, and spin precession will produce a richer frequency structure.  Finally, incorporating radiation reaction, gravitational self-force effects, and more realistic emission prescriptions would be necessary for precision applications to gravitational-wave or X-ray timing observations.

\acknowledgments
A. \"O. and R. P. would like to acknowledge networking support of the COST Action CA21106 - COSMIC WISPers in the Dark Universe: Theory, astrophysics and experiments (CosmicWISPers), the COST Action CA22113 - Fundamental challenges in theoretical physics (THEORY-CHALLENGES), the COST Action CA21136 - Addressing observational tensions in cosmology with systematics and fundamental physics (CosmoVerse), the COST Action CA23130 - Bridging high and low energies in search of quantum gravity (BridgeQG), and the COST Action CA23115 - Relativistic Quantum Information (RQI) funded by COST (European Cooperation in Science and Technology). A. \"O. would also like to acknowledge the funding support of SCOAP3, Switzerland and TUBITAK, Turkiye.

\appendix

\section{The full expressions of the specific angular momentum $l$ and  specific energy $\mathcal{E}$ for circular orbits}\label{apdxA}
    \begin{subequations}
        \begin{align}
            &\mathcal{E}_c=\sqrt{\frac{\Phi+\Psi-\Xi+\Theta}{\mathcal{A}}} -sr \sqrt{\frac{\mathcal{B}}{\mathcal{A}}}+3 M s \sqrt{\frac{\mathcal{D}+\mathcal{F}}{\mathcal{A}}}\,,\\
            &l_c=\sqrt{\frac{r^6 \left(\mathcal{G}+\mathcal{I}\right)}{\mathcal{A}}}\,,
        \end{align}
    \end{subequations}
in which
\begin{widetext}
    \begin{subequations}
        \begin{align}
& \Phi=-972 s^4 M^5+27 \left(65 r s^4+6 r^3 s^2\right) M^4-3 \left(8 r^6+88 s^2 r^4+402 s^4 r^2+27 s^3 \sqrt{4 M (r-2 M)^2 r^3+(r-4 M)^2 (3 M-2 r)^2 s^2}\right) M^3\,\\
&\Psi=\Big(32 r^7+159 s^2 r^5+3 \left(133 s^4+6 \sqrt{4 M (r-2 M)^2 r^3+(r-4 M)^2 (3 M-2 r)^2 s^2} s\right) r^3\\\nonumber
&+72 s^3 \sqrt{4 M (r-2 M)^2 r^3+(r-4 M)^2 (3 M-2 r)^2 s^2} r\Big) M^2\,,\\
&\Xi=Mr^2\Big[14 r^6+42 s^2 r^4+64s^4r^2+s\sqrt{4 M (r-2 M)^2 r^3+(r-4 M)^2 (3 M-2 r)^2 s^2}\left(13r^2+21s^2\right)\Big]\,,\\
&\Theta=2 r^3\left[r^6+2 s^2 r^4+2s^4r^2+s\sqrt{4 M (r-2 M)^2 r^3+(r-4 M)^2 (3 M-2 r)^2 s^2}\left( r^2+s^2\right)\right]\,,\\
&\mathcal{A}=2\left[(r-3 M)^2 r^7+M (2 r-9 M) s^2 r^5\right]\,,\\
&\mathcal{B}=-108 s^2 M^3+\left(123 r s^2-6 r^3\right) M^2+\left(2 r^4-40 s^2 r^2-9 s \sqrt{4 M (r-2 M)^2 r^3+(r-4 M)^2 (3 M-2 r)^2 s^2}\right)M\\\nonumber
&+2 r s \left(2 s r^2+\sqrt{4 M (r-2 M)^2 r^3+(r-4 M)^2 (3 M-2 r)^2 s^2}\right)\,,\\
&\mathcal{D}=-108 s^2 M^3+\left(123 r s^2-6 r^3\right) M^2+\left(2 r^4-40 s^2 r^2-9 s \sqrt{4 M (r-2 M)^2 r^3+(r-4 M)^2 (3 M-2 r)^2 s^2}\right) M\,,\\
&\mathcal{F}=2 r s \left(2 s r^2+\sqrt{4 M (r-2 M)^2 r^3+(r-4 M)^2 (3 M-2 r)^2 s^2}\right)\,,\\
&\mathcal{G}=-108 M^3 s^2+M^2 \left(123 r s^2-6 r^3\right)+2 r s \left(\sqrt{4 M r^3 (r-2 M)^2+s^2 (r-4 M)^2 (3 M-2 r)^2}+2 r^2 s\right)\,,\\
&\mathcal{I}=M \left(-9 s \sqrt{4 M r^3 (r-2 M)^2+s^2 (r-4 M)^2 (3 M-2 r)^2}+2 r^4-40 r^2 s^2\right)\,,
        \end{align}
    \end{subequations}
\end{widetext}

\section{The full expressions of the specific angular momentum $l_p$ and specific energy $\mathcal{E}_p$}\label{apdx2}
\begin{subequations}
\begin{align}
&\mathcal{E}_p=\frac{\left(\Phi_1+\sqrt{\Psi_1+\Xi_1}\right)}{2 \lambda ^3 M}l_p\,,\\
&l_p=\sqrt{\frac{\mathcal{H}}{2\mathcal{A}_1}}\,,
\end{align}
\end{subequations}
in which 
\begin{widetext}
    \begin{subequations}
        \begin{align}
            &\Phi_1=12 \left(e^2-1\right) M^2 s+\left(e^2+11\right) \lambda  M s-2 \lambda ^2 s\,,\\
            &\Psi_1=144 \left(e^2-1\right)^2 M^4 s^2+8 \left(e^2-1\right) \lambda  M^3 \left[3 \left(e^2+11\right) s^2-2 \lambda ^2\right]+\lambda ^2 M^2 \left[\left(e^2-13\right)^2 s^2-16 \lambda ^2\right]\,,\\
            &\Xi_1=4 \lambda ^3 M \left[\lambda ^2-\left(e^2+11\right) s^2\right]+4 \lambda ^4 s^2\,,\\
            &\mathcal{H}=\lambda  \Big[-M \left(3 \left[e^2+3\right] s \sqrt{\Theta_1}+8 \left(e^2+5\right) \lambda ^2 s^2-2 \lambda ^4\right)+2 \lambda  s \left(\sqrt{\Theta_1}+2 \lambda ^2 s\right)+36 \left(e^4+2 e^2-3\right) M^3 s^2\\\nonumber
            &+\lambda  M^2 \left(3 \left[e^4+6 e^2+41\right] s^2-2 \left(e^2+3\right) \lambda ^2\right)\Big]\,,\\
            &\Theta_1=144 \left(e^2-1\right)^2 M^4 s^2+8 \left(e^2-1\right) \lambda  M^3 \left[3 \left(e^2+11\right) s^2-2 \lambda ^2\right]+\lambda ^2 M^2 \left(\left(e^2-13\right)^2 s^2-16 \lambda ^2\right)\\\nonumber
            &+4 \lambda ^3 M \left[\lambda ^2-\left(e^2+11\right) s^2\right]+4 \lambda ^4 s^2\,,\\
            &\mathcal{A}_1=\left(e^2+3\right) M^2 \left[\left(e^2+3\right) \lambda ^2-3 \left(e^2-1\right)^2 s^2\right]-2 \left(e^2+3\right) \lambda ^3 M+2 \left(e^2-1\right)^2 \lambda  M s^2+\lambda ^4
        \end{align}
    \end{subequations}
\end{widetext}

\bibliography{ref}

\begin{thebibliography}{44}%
\makeatletter
\providecommand \@ifxundefined [1]{%
 \@ifx{#1\undefined}
}%
\providecommand \@ifnum [1]{%
 \ifnum #1\expandafter \@firstoftwo
 \else \expandafter \@secondoftwo
 \fi
}%
\providecommand \@ifx [1]{%
 \ifx #1\expandafter \@firstoftwo
 \else \expandafter \@secondoftwo
 \fi
}%
\providecommand \natexlab [1]{#1}%
\providecommand \enquote  [1]{``#1''}%
\providecommand \bibnamefont  [1]{#1}%
\providecommand \bibfnamefont [1]{#1}%
\providecommand \citenamefont [1]{#1}%
\providecommand \href@noop [0]{\@secondoftwo}%
\providecommand \href [0]{\begingroup \@sanitize@url \@href}%
\providecommand \@href[1]{\@@startlink{#1}\@@href}%
\providecommand \@@href[1]{\endgroup#1\@@endlink}%
\providecommand \@sanitize@url [0]{\catcode `\\12\catcode `\$12\catcode `\&12\catcode `\#12\catcode `\^12\catcode `\_12\catcode `\%12\relax}%
\providecommand \@@startlink[1]{}%
\providecommand \@@endlink[0]{}%
\providecommand \url  [0]{\begingroup\@sanitize@url \@url }%
\providecommand \@url [1]{\endgroup\@href {#1}{\urlprefix }}%
\providecommand \urlprefix  [0]{URL }%
\providecommand \Eprint [0]{\href }%
\providecommand \doibase [0]{https://doi.org/}%
\providecommand \selectlanguage [0]{\@gobble}%
\providecommand \bibinfo  [0]{\@secondoftwo}%
\providecommand \bibfield  [0]{\@secondoftwo}%
\providecommand \translation [1]{[#1]}%
\providecommand \BibitemOpen [0]{}%
\providecommand \bibitemStop [0]{}%
\providecommand \bibitemNoStop [0]{.\EOS\space}%
\providecommand \EOS [0]{\spacefactor3000\relax}%
\providecommand \BibitemShut  [1]{\csname bibitem#1\endcsname}%
\let\auto@bib@innerbib\@empty
\bibitem [{\citenamefont {Berti}\ \emph {et~al.}(2015)\citenamefont {Berti} \emph {et~al.}}]{Berti:2015itd}%
  \BibitemOpen
  \bibfield  {author} {\bibinfo {author} {\bibfnamefont {E.}~\bibnamefont {Berti}} \emph {et~al.},\ }\bibfield  {title} {\bibinfo {title} {{Testing General Relativity with Present and Future Astrophysical Observations}},\ }\href {https://doi.org/10.1088/0264-9381/32/24/243001} {\bibfield  {journal} {\bibinfo  {journal} {Class. Quant. Grav.}\ }\textbf {\bibinfo {volume} {32}},\ \bibinfo {pages} {243001} (\bibinfo {year} {2015})},\ \Eprint {https://arxiv.org/abs/1501.07274} {arXiv:1501.07274 [gr-qc]} \BibitemShut {NoStop}%
\bibitem [{\citenamefont {Cardoso}\ and\ \citenamefont {Pani}(2019)}]{Cardoso:2019rvt}%
  \BibitemOpen
  \bibfield  {author} {\bibinfo {author} {\bibfnamefont {V.}~\bibnamefont {Cardoso}}\ and\ \bibinfo {author} {\bibfnamefont {P.}~\bibnamefont {Pani}},\ }\bibfield  {title} {\bibinfo {title} {{Testing the nature of dark compact objects: a status report}},\ }\href {https://doi.org/10.1007/s41114-019-0020-4} {\bibfield  {journal} {\bibinfo  {journal} {Living Rev. Rel.}\ }\textbf {\bibinfo {volume} {22}},\ \bibinfo {pages} {4} (\bibinfo {year} {2019})},\ \Eprint {https://arxiv.org/abs/1904.05363} {arXiv:1904.05363 [gr-qc]} \BibitemShut {NoStop}%
\bibitem [{\citenamefont {Vagnozzi}\ \emph {et~al.}(2023)\citenamefont {Vagnozzi} \emph {et~al.}}]{Vagnozzi:2022moj}%
  \BibitemOpen
  \bibfield  {author} {\bibinfo {author} {\bibfnamefont {S.}~\bibnamefont {Vagnozzi}} \emph {et~al.},\ }\bibfield  {title} {\bibinfo {title} {{Horizon-scale tests of gravity theories and fundamental physics from the Event Horizon Telescope image of Sagittarius A}},\ }\href {https://doi.org/10.1088/1361-6382/acd97b} {\bibfield  {journal} {\bibinfo  {journal} {Class. Quant. Grav.}\ }\textbf {\bibinfo {volume} {40}},\ \bibinfo {pages} {165007} (\bibinfo {year} {2023})},\ \Eprint {https://arxiv.org/abs/2205.07787} {arXiv:2205.07787 [gr-qc]} \BibitemShut {NoStop}%
\bibitem [{\citenamefont {Battista}(2024)}]{Battista:2023iyu}%
  \BibitemOpen
  \bibfield  {author} {\bibinfo {author} {\bibfnamefont {E.}~\bibnamefont {Battista}},\ }\bibfield  {title} {\bibinfo {title} {{Quantum Schwarzschild geometry in effective field theory models of gravity}},\ }\href {https://doi.org/10.1103/PhysRevD.109.026004} {\bibfield  {journal} {\bibinfo  {journal} {Phys. Rev. D}\ }\textbf {\bibinfo {volume} {109}},\ \bibinfo {pages} {026004} (\bibinfo {year} {2024})},\ \Eprint {https://arxiv.org/abs/2312.00450} {arXiv:2312.00450 [gr-qc]} \BibitemShut {NoStop}%
\bibitem [{\citenamefont {Battista}\ \emph {et~al.}(2026)\citenamefont {Battista}, \citenamefont {Capozziello},\ and\ \citenamefont {Chen}}]{Battista:2026nsx}%
  \BibitemOpen
  \bibfield  {author} {\bibinfo {author} {\bibfnamefont {E.}~\bibnamefont {Battista}}, \bibinfo {author} {\bibfnamefont {S.}~\bibnamefont {Capozziello}},\ and\ \bibinfo {author} {\bibfnamefont {C.-Y.}\ \bibnamefont {Chen}},\ }\bibfield  {title} {\bibinfo {title} {{Shadow signatures and energy accumulation in Lorentzian-Euclidean black holes}},\ }\href {https://doi.org/10.1103/zf2w-7fqn} {\bibfield  {journal} {\bibinfo  {journal} {Phys. Rev. D}\ }\textbf {\bibinfo {volume} {113}},\ \bibinfo {pages} {104039} (\bibinfo {year} {2026})},\ \Eprint {https://arxiv.org/abs/2601.10806} {arXiv:2601.10806 [gr-qc]} \BibitemShut {NoStop}%
\bibitem [{\citenamefont {Mathisson}(1937)}]{Mathisson:1937NM}%
  \BibitemOpen
  \bibfield  {author} {\bibinfo {author} {\bibfnamefont {M.}~\bibnamefont {Mathisson}},\ }\bibfield  {title} {\bibinfo {title} {Neue mechanik materieller systeme},\ }\href@noop {} {\bibfield  {journal} {\bibinfo  {journal} {Acta Physica Polonica}\ }\textbf {\bibinfo {volume} {6}},\ \bibinfo {pages} {163} (\bibinfo {year} {1937})}\BibitemShut {NoStop}%
\bibitem [{\citenamefont {Papapetrou}(1951)}]{Papapetrou:1951}%
  \BibitemOpen
  \bibfield  {author} {\bibinfo {author} {\bibfnamefont {A.}~\bibnamefont {Papapetrou}},\ }\bibfield  {title} {\bibinfo {title} {Spinning test-particles in general relativity. i},\ }\href {https://doi.org/10.1098/rspa.1951.0200} {\bibfield  {journal} {\bibinfo  {journal} {Proceedings of the Royal Society of London. Series A, Mathematical and Physical Sciences}\ }\textbf {\bibinfo {volume} {209}},\ \bibinfo {pages} {248} (\bibinfo {year} {1951})}\BibitemShut {NoStop}%
\bibitem [{\citenamefont {Tulczyjew}(1959)}]{Tulczyjew:1959}%
  \BibitemOpen
  \bibfield  {author} {\bibinfo {author} {\bibfnamefont {W.~M.}\ \bibnamefont {Tulczyjew}},\ }\bibfield  {title} {\bibinfo {title} {Motion of multipole particles in general relativity theory},\ }\href@noop {} {\bibfield  {journal} {\bibinfo  {journal} {Acta Physica Polonica}\ }\textbf {\bibinfo {volume} {18}},\ \bibinfo {pages} {393} (\bibinfo {year} {1959})}\BibitemShut {NoStop}%
\bibitem [{\citenamefont {Dixon}(1964)}]{Dixon:1964}%
  \BibitemOpen
  \bibfield  {author} {\bibinfo {author} {\bibfnamefont {W.~G.}\ \bibnamefont {Dixon}},\ }\bibfield  {title} {\bibinfo {title} {A covariant multipole formalism for extended test bodies in general relativity},\ }\href {https://doi.org/10.1007/BF02734579} {\bibfield  {journal} {\bibinfo  {journal} {Il Nuovo Cimento}\ }\textbf {\bibinfo {volume} {34}},\ \bibinfo {pages} {317} (\bibinfo {year} {1964})}\BibitemShut {NoStop}%
\bibitem [{\citenamefont {Costa}\ and\ \citenamefont {Nat{\'a}rio}(2015)}]{Costa:2014nta}%
  \BibitemOpen
  \bibfield  {author} {\bibinfo {author} {\bibfnamefont {L.~F.~O.}\ \bibnamefont {Costa}}\ and\ \bibinfo {author} {\bibfnamefont {J.}~\bibnamefont {Nat{\'a}rio}},\ }\bibfield  {title} {\bibinfo {title} {{Center of mass, spin supplementary conditions, and the momentum of spinning particles}},\ }\href {https://doi.org/10.1007/978-3-319-18335-0_6} {\bibfield  {journal} {\bibinfo  {journal} {Fund. Theor. Phys.}\ }\textbf {\bibinfo {volume} {179}},\ \bibinfo {pages} {215} (\bibinfo {year} {2015})},\ \Eprint {https://arxiv.org/abs/1410.6443} {arXiv:1410.6443 [gr-qc]} \BibitemShut {NoStop}%
\bibitem [{\citenamefont {Semerak}(1999)}]{Semerak:1999qc}%
  \BibitemOpen
  \bibfield  {author} {\bibinfo {author} {\bibfnamefont {O.}~\bibnamefont {Semerak}},\ }\bibfield  {title} {\bibinfo {title} {{Spinning test particles in a Kerr field. 1.}},\ }\href {https://doi.org/10.1046/j.1365-8711.1999.02754.x} {\bibfield  {journal} {\bibinfo  {journal} {Mon. Not. Roy. Astron. Soc.}\ }\textbf {\bibinfo {volume} {308}},\ \bibinfo {pages} {863} (\bibinfo {year} {1999})}\BibitemShut {NoStop}%
\bibitem [{\citenamefont {Kyrian}\ and\ \citenamefont {Semerak}(2007)}]{Kyrian:2007zz}%
  \BibitemOpen
  \bibfield  {author} {\bibinfo {author} {\bibfnamefont {K.}~\bibnamefont {Kyrian}}\ and\ \bibinfo {author} {\bibfnamefont {O.}~\bibnamefont {Semerak}},\ }\bibfield  {title} {\bibinfo {title} {{Spinning test particles in a Kerr field}},\ }\href {https://doi.org/10.1111/j.1365-2966.2007.12502.x} {\bibfield  {journal} {\bibinfo  {journal} {Mon. Not. Roy. Astron. Soc.}\ }\textbf {\bibinfo {volume} {382}},\ \bibinfo {pages} {1922} (\bibinfo {year} {2007})}\BibitemShut {NoStop}%
\bibitem [{\citenamefont {Stella}\ and\ \citenamefont {Vietri}(1998)}]{Stella:1998lense}%
  \BibitemOpen
  \bibfield  {author} {\bibinfo {author} {\bibfnamefont {L.}~\bibnamefont {Stella}}\ and\ \bibinfo {author} {\bibfnamefont {M.}~\bibnamefont {Vietri}},\ }\bibfield  {title} {\bibinfo {title} {Lense-thirring precession and quasi-periodic oscillations in low-mass x-ray binaries},\ }\href {https://doi.org/10.1086/311075} {\bibfield  {journal} {\bibinfo  {journal} {Astrophysical Journal Letters}\ }\textbf {\bibinfo {volume} {492}},\ \bibinfo {pages} {L59} (\bibinfo {year} {1998})},\ \Eprint {https://arxiv.org/abs/astro-ph/9709085} {arXiv:astro-ph/9709085} \BibitemShut {NoStop}%
\bibitem [{\citenamefont {Stella}\ and\ \citenamefont {Vietri}(1999)}]{Stella:1999prl}%
  \BibitemOpen
  \bibfield  {author} {\bibinfo {author} {\bibfnamefont {L.}~\bibnamefont {Stella}}\ and\ \bibinfo {author} {\bibfnamefont {M.}~\bibnamefont {Vietri}},\ }\bibfield  {title} {\bibinfo {title} {khz quasiperiodic oscillations in low-mass x-ray binaries as probes of general relativity in the strong-field regime},\ }\href {https://doi.org/10.1103/PhysRevLett.82.17} {\bibfield  {journal} {\bibinfo  {journal} {Physical Review Letters}\ }\textbf {\bibinfo {volume} {82}},\ \bibinfo {pages} {17} (\bibinfo {year} {1999})},\ \Eprint {https://arxiv.org/abs/astro-ph/9812124} {arXiv:astro-ph/9812124} \BibitemShut {NoStop}%
\bibitem [{\citenamefont {Klu{\'z}niak}\ and\ \citenamefont {Abramowicz}(2001)}]{Kluzniak:2001ar}%
  \BibitemOpen
  \bibfield  {author} {\bibinfo {author} {\bibfnamefont {W.}~\bibnamefont {Klu{\'z}niak}}\ and\ \bibinfo {author} {\bibfnamefont {M.~A.}\ \bibnamefont {Abramowicz}},\ }\href@noop {} {\bibinfo {title} {The physics of khz qpos---strong gravity's coupled anharmonic oscillators}} (\bibinfo {year} {2001}),\ \Eprint {https://arxiv.org/abs/astro-ph/0105057} {arXiv:astro-ph/0105057} \BibitemShut {NoStop}%
\bibitem [{\citenamefont {Abramowicz}\ and\ \citenamefont {Klu{\'z}niak}(2001)}]{Abramowicz:2001bi}%
  \BibitemOpen
  \bibfield  {author} {\bibinfo {author} {\bibfnamefont {M.~A.}\ \bibnamefont {Abramowicz}}\ and\ \bibinfo {author} {\bibfnamefont {W.}~\bibnamefont {Klu{\'z}niak}},\ }\bibfield  {title} {\bibinfo {title} {A precise determination of black hole spin in gro j1655-40},\ }\href {https://doi.org/10.1051/0004-6361:20010791} {\bibfield  {journal} {\bibinfo  {journal} {Astronomy \& Astrophysics}\ }\textbf {\bibinfo {volume} {374}},\ \bibinfo {pages} {L19} (\bibinfo {year} {2001})},\ \Eprint {https://arxiv.org/abs/astro-ph/0105077} {arXiv:astro-ph/0105077} \BibitemShut {NoStop}%
\bibitem [{\citenamefont {Remillard}\ and\ \citenamefont {McClintock}(2006)}]{Remillard:2006fc}%
  \BibitemOpen
  \bibfield  {author} {\bibinfo {author} {\bibfnamefont {R.~A.}\ \bibnamefont {Remillard}}\ and\ \bibinfo {author} {\bibfnamefont {J.~E.}\ \bibnamefont {McClintock}},\ }\bibfield  {title} {\bibinfo {title} {{X-ray Properties of Black-Hole Binaries}},\ }\href {https://doi.org/10.1146/annurev.astro.44.051905.092532} {\bibfield  {journal} {\bibinfo  {journal} {Ann. Rev. Astron. Astrophys.}\ }\textbf {\bibinfo {volume} {44}},\ \bibinfo {pages} {49} (\bibinfo {year} {2006})},\ \Eprint {https://arxiv.org/abs/astro-ph/0606352} {arXiv:astro-ph/0606352} \BibitemShut {NoStop}%
\bibitem [{\citenamefont {Abramowicz}\ and\ \citenamefont {Fragile}(2013)}]{Abramowicz:2011xu}%
  \BibitemOpen
  \bibfield  {author} {\bibinfo {author} {\bibfnamefont {M.~A.}\ \bibnamefont {Abramowicz}}\ and\ \bibinfo {author} {\bibfnamefont {P.~C.}\ \bibnamefont {Fragile}},\ }\bibfield  {title} {\bibinfo {title} {{Foundations of Black Hole Accretion Disk Theory}},\ }\href {https://doi.org/10.12942/lrr-2013-1} {\bibfield  {journal} {\bibinfo  {journal} {Living Rev. Rel.}\ }\textbf {\bibinfo {volume} {16}},\ \bibinfo {pages} {1} (\bibinfo {year} {2013})},\ \Eprint {https://arxiv.org/abs/1104.5499} {arXiv:1104.5499 [astro-ph.HE]} \BibitemShut {NoStop}%
\bibitem [{\citenamefont {Ingram}\ and\ \citenamefont {Motta}(2019)}]{Ingram:2019mna}%
  \BibitemOpen
  \bibfield  {author} {\bibinfo {author} {\bibfnamefont {A.}~\bibnamefont {Ingram}}\ and\ \bibinfo {author} {\bibfnamefont {S.}~\bibnamefont {Motta}},\ }\bibfield  {title} {\bibinfo {title} {{A review of quasi-periodic oscillations from black hole X-ray binaries: observation and theory}},\ }\href {https://doi.org/10.1016/j.newar.2020.101524} {\bibfield  {journal} {\bibinfo  {journal} {New Astron. Rev.}\ }\textbf {\bibinfo {volume} {85}},\ \bibinfo {pages} {101524} (\bibinfo {year} {2019})},\ \Eprint {https://arxiv.org/abs/2001.08758} {arXiv:2001.08758 [astro-ph.HE]} \BibitemShut {NoStop}%
\bibitem [{\citenamefont {Chandrasekhar}(1983)}]{Chandrasekhar:1983}%
  \BibitemOpen
  \bibfield  {author} {\bibinfo {author} {\bibfnamefont {S.}~\bibnamefont {Chandrasekhar}},\ }\href@noop {} {\emph {\bibinfo {title} {The Mathematical Theory of Black Holes}}}\ (\bibinfo  {publisher} {Clarendon Press},\ \bibinfo {address} {Oxford},\ \bibinfo {year} {1983})\BibitemShut {NoStop}%
\bibitem [{\citenamefont {Witzany}\ and\ \citenamefont {Piovano}(2024)}]{Witzany:2023bmq}%
  \BibitemOpen
  \bibfield  {author} {\bibinfo {author} {\bibfnamefont {V.}~\bibnamefont {Witzany}}\ and\ \bibinfo {author} {\bibfnamefont {G.~A.}\ \bibnamefont {Piovano}},\ }\bibfield  {title} {\bibinfo {title} {Analytic solutions for the motion of spinning particles near spherically symmetric black holes and exotic compact objects},\ }\href {https://doi.org/10.1103/PhysRevLett.132.171401} {\bibfield  {journal} {\bibinfo  {journal} {Physical Review Letters}\ }\textbf {\bibinfo {volume} {132}},\ \bibinfo {pages} {171401} (\bibinfo {year} {2024})},\ \Eprint {https://arxiv.org/abs/2308.00021} {arXiv:2308.00021 [gr-qc]} \BibitemShut {NoStop}%
\bibitem [{\citenamefont {Suzuki}\ and\ \citenamefont {Maeda}(1997)}]{Suzuki:1996gm}%
  \BibitemOpen
  \bibfield  {author} {\bibinfo {author} {\bibfnamefont {S.}~\bibnamefont {Suzuki}}\ and\ \bibinfo {author} {\bibfnamefont {K.-i.}\ \bibnamefont {Maeda}},\ }\bibfield  {title} {\bibinfo {title} {{Chaos in Schwarzschild space-time: The motion of a spinning particle}},\ }\href {https://doi.org/10.1103/PhysRevD.55.4848} {\bibfield  {journal} {\bibinfo  {journal} {Phys. Rev. D}\ }\textbf {\bibinfo {volume} {55}},\ \bibinfo {pages} {4848} (\bibinfo {year} {1997})},\ \Eprint {https://arxiv.org/abs/gr-qc/9604020} {arXiv:gr-qc/9604020} \BibitemShut {NoStop}%
\bibitem [{\citenamefont {Jefremov}\ \emph {et~al.}(2015)\citenamefont {Jefremov}, \citenamefont {Tsupko},\ and\ \citenamefont {Bisnovatyi-Kogan}}]{Jefremov:2015gza}%
  \BibitemOpen
  \bibfield  {author} {\bibinfo {author} {\bibfnamefont {P.~I.}\ \bibnamefont {Jefremov}}, \bibinfo {author} {\bibfnamefont {O.~Y.}\ \bibnamefont {Tsupko}},\ and\ \bibinfo {author} {\bibfnamefont {G.~S.}\ \bibnamefont {Bisnovatyi-Kogan}},\ }\bibfield  {title} {\bibinfo {title} {{Innermost stable circular orbits of spinning test particles in Schwarzschild and Kerr space-times}},\ }\href {https://doi.org/10.1103/PhysRevD.91.124030} {\bibfield  {journal} {\bibinfo  {journal} {Phys. Rev. D}\ }\textbf {\bibinfo {volume} {91}},\ \bibinfo {pages} {124030} (\bibinfo {year} {2015})},\ \Eprint {https://arxiv.org/abs/1503.07060} {arXiv:1503.07060 [gr-qc]} \BibitemShut {NoStop}%
\bibitem [{\citenamefont {Levin}\ and\ \citenamefont {Perez-Giz}(2008)}]{Levin:2008mq}%
  \BibitemOpen
  \bibfield  {author} {\bibinfo {author} {\bibfnamefont {J.}~\bibnamefont {Levin}}\ and\ \bibinfo {author} {\bibfnamefont {G.}~\bibnamefont {Perez-Giz}},\ }\bibfield  {title} {\bibinfo {title} {A periodic table for black hole orbits},\ }\href {https://doi.org/10.1103/PhysRevD.77.103005} {\bibfield  {journal} {\bibinfo  {journal} {Physical Review D}\ }\textbf {\bibinfo {volume} {77}},\ \bibinfo {pages} {103005} (\bibinfo {year} {2008})},\ \Eprint {https://arxiv.org/abs/0802.0459} {arXiv:0802.0459 [gr-qc]} \BibitemShut {NoStop}%
\bibitem [{\citenamefont {Alloqulov}\ \emph {et~al.}(2026{\natexlab{a}})\citenamefont {Alloqulov}, \citenamefont {Shaymatov}, \citenamefont {Ahmedov},\ and\ \citenamefont {Zhu}}]{Alloqulov:2026regularBH}%
  \BibitemOpen
  \bibfield  {author} {\bibinfo {author} {\bibfnamefont {M.}~\bibnamefont {Alloqulov}}, \bibinfo {author} {\bibfnamefont {S.}~\bibnamefont {Shaymatov}}, \bibinfo {author} {\bibfnamefont {B.}~\bibnamefont {Ahmedov}},\ and\ \bibinfo {author} {\bibfnamefont {T.}~\bibnamefont {Zhu}},\ }\bibfield  {title} {\bibinfo {title} {Regular black hole's impact on the gravitational waveforms from periodic orbits},\ }\href {https://doi.org/10.1140/epjc/s10052-025-15251-1} {\bibfield  {journal} {\bibinfo  {journal} {European Physical Journal C}\ }\textbf {\bibinfo {volume} {86}},\ \bibinfo {pages} {117} (\bibinfo {year} {2026}{\natexlab{a}})},\ \Eprint {https://arxiv.org/abs/2508.05245} {arXiv:2508.05245 [gr-qc]} \BibitemShut {NoStop}%
\bibitem [{\citenamefont {Yang}\ \emph {et~al.}(2025)\citenamefont {Yang}, \citenamefont {Zhang}, \citenamefont {Zhu}, \citenamefont {Zhao},\ and\ \citenamefont {Liu}}]{Yang:2025qcbh}%
  \BibitemOpen
  \bibfield  {author} {\bibinfo {author} {\bibfnamefont {S.}~\bibnamefont {Yang}}, \bibinfo {author} {\bibfnamefont {Y.-P.}\ \bibnamefont {Zhang}}, \bibinfo {author} {\bibfnamefont {T.}~\bibnamefont {Zhu}}, \bibinfo {author} {\bibfnamefont {L.}~\bibnamefont {Zhao}},\ and\ \bibinfo {author} {\bibfnamefont {Y.-X.}\ \bibnamefont {Liu}},\ }\bibfield  {title} {\bibinfo {title} {Gravitational waveforms from periodic orbits around a quantum-corrected black hole},\ }\href {https://doi.org/10.1088/1475-7516/2025/01/091} {\bibfield  {journal} {\bibinfo  {journal} {Journal of Cosmology and Astroparticle Physics}\ }\textbf {\bibinfo {volume} {2025}}\bibfield  {number} {\bibinfo  {number} { (01)},\ \bibinfo {pages} {091}},\ }\Eprint {https://arxiv.org/abs/2407.00283} {arXiv:2407.00283 [gr-qc]} \BibitemShut {NoStop}%
\bibitem [{\citenamefont {Babak}\ \emph {et~al.}(2007)\citenamefont {Babak}, \citenamefont {Fang}, \citenamefont {Gair}, \citenamefont {Glampedakis},\ and\ \citenamefont {Hughes}}]{Babak:2006uv}%
  \BibitemOpen
  \bibfield  {author} {\bibinfo {author} {\bibfnamefont {S.}~\bibnamefont {Babak}}, \bibinfo {author} {\bibfnamefont {H.}~\bibnamefont {Fang}}, \bibinfo {author} {\bibfnamefont {J.~R.}\ \bibnamefont {Gair}}, \bibinfo {author} {\bibfnamefont {K.}~\bibnamefont {Glampedakis}},\ and\ \bibinfo {author} {\bibfnamefont {S.~A.}\ \bibnamefont {Hughes}},\ }\bibfield  {title} {\bibinfo {title} {{'Kludge' gravitational waveforms for a test-body orbiting a Kerr black hole}},\ }\href {https://doi.org/10.1103/PhysRevD.75.024005} {\bibfield  {journal} {\bibinfo  {journal} {Phys. Rev. D}\ }\textbf {\bibinfo {volume} {75}},\ \bibinfo {pages} {024005} (\bibinfo {year} {2007})},\ \bibinfo {note} {[Erratum: Phys.Rev.D 77, 04990 (2008)]},\ \Eprint {https://arxiv.org/abs/gr-qc/0607007} {arXiv:gr-qc/0607007} \BibitemShut {NoStop}%
\bibitem [{\citenamefont {Khan}\ \emph {et~al.}(2024)\citenamefont {Khan}, \citenamefont {Uktamov}, \citenamefont {Rayimbaev}, \citenamefont {Abdujabbarov}, \citenamefont {Ibragimov},\ and\ \citenamefont {Chen}}]{Khan:2024jez}%
  \BibitemOpen
  \bibfield  {author} {\bibinfo {author} {\bibfnamefont {S.~U.}\ \bibnamefont {Khan}}, \bibinfo {author} {\bibfnamefont {U.}~\bibnamefont {Uktamov}}, \bibinfo {author} {\bibfnamefont {J.}~\bibnamefont {Rayimbaev}}, \bibinfo {author} {\bibfnamefont {A.}~\bibnamefont {Abdujabbarov}}, \bibinfo {author} {\bibfnamefont {I.}~\bibnamefont {Ibragimov}},\ and\ \bibinfo {author} {\bibfnamefont {Z.-M.}\ \bibnamefont {Chen}},\ }\bibfield  {title} {\bibinfo {title} {Circular orbits and collisions of particles with magnetic dipole moment near magnetized kerr black holes in modified gravity},\ }\href {https://doi.org/10.1140/epjc/s10052-024-12567-2} {\bibfield  {journal} {\bibinfo  {journal} {European Physical Journal C}\ }\textbf {\bibinfo {volume} {84}},\ \bibinfo {pages} {203} (\bibinfo {year} {2024})}\BibitemShut {NoStop}%
\bibitem [{\citenamefont {Uktamov}\ \emph {et~al.}(2024)\citenamefont {Uktamov}, \citenamefont {Fathi}, \citenamefont {Rayimbaev},\ and\ \citenamefont {Abdujabbarov}}]{Uktamov:2024magdip}%
  \BibitemOpen
  \bibfield  {author} {\bibinfo {author} {\bibfnamefont {U.}~\bibnamefont {Uktamov}}, \bibinfo {author} {\bibfnamefont {M.}~\bibnamefont {Fathi}}, \bibinfo {author} {\bibfnamefont {J.}~\bibnamefont {Rayimbaev}},\ and\ \bibinfo {author} {\bibfnamefont {A.}~\bibnamefont {Abdujabbarov}},\ }\bibfield  {title} {\bibinfo {title} {Orbits of particles with magnetic dipole moment around magnetized schwarzschild black holes: Applications to the s2 star orbit},\ }\href {https://doi.org/10.1103/PhysRevD.110.084084} {\bibfield  {journal} {\bibinfo  {journal} {Physical Review D}\ }\textbf {\bibinfo {volume} {110}},\ \bibinfo {pages} {084084} (\bibinfo {year} {2024})},\ \Eprint {https://arxiv.org/abs/2406.03371} {arXiv:2406.03371 [gr-qc]} \BibitemShut {NoStop}%
\bibitem [{\citenamefont {Uktamov}\ \emph {et~al.}(2025{\natexlab{a}})\citenamefont {Uktamov}, \citenamefont {Shaymatov}, \citenamefont {Ahmedov},\ and\ \citenamefont {Yuan}}]{Uktamov:2025dmhalo}%
  \BibitemOpen
  \bibfield  {author} {\bibinfo {author} {\bibfnamefont {U.}~\bibnamefont {Uktamov}}, \bibinfo {author} {\bibfnamefont {S.}~\bibnamefont {Shaymatov}}, \bibinfo {author} {\bibfnamefont {B.}~\bibnamefont {Ahmedov}},\ and\ \bibinfo {author} {\bibfnamefont {C.}~\bibnamefont {Yuan}},\ }\bibfield  {title} {\bibinfo {title} {New analytical model of static black hole with a dark matter halo and parametric constraints through quasiperiodic oscillations},\ }\href {https://doi.org/10.1140/epjc/s10052-025-15171-0} {\bibfield  {journal} {\bibinfo  {journal} {European Physical Journal C}\ }\textbf {\bibinfo {volume} {85}},\ \bibinfo {pages} {1432} (\bibinfo {year} {2025}{\natexlab{a}})}\BibitemShut {NoStop}%
\bibitem [{\citenamefont {Uktamov}\ \emph {et~al.}(2025{\natexlab{b}})\citenamefont {Uktamov}, \citenamefont {Narzilloev}, \citenamefont {Abdujabbarov},\ and\ \citenamefont {Ahmedov}}]{Uktamov:2025stringy}%
  \BibitemOpen
  \bibfield  {author} {\bibinfo {author} {\bibfnamefont {U.}~\bibnamefont {Uktamov}}, \bibinfo {author} {\bibfnamefont {B.}~\bibnamefont {Narzilloev}}, \bibinfo {author} {\bibfnamefont {A.}~\bibnamefont {Abdujabbarov}},\ and\ \bibinfo {author} {\bibfnamefont {B.}~\bibnamefont {Ahmedov}},\ }\bibfield  {title} {\bibinfo {title} {Electrically charged stringy black hole versus magnetically charged stringy black hole in terms of optical properties},\ }\href {https://doi.org/10.1142/S0218271825500774} {\bibfield  {journal} {\bibinfo  {journal} {International Journal of Modern Physics D}\ }\textbf {\bibinfo {volume} {34}},\ \bibinfo {pages} {2550077} (\bibinfo {year} {2025}{\natexlab{b}})}\BibitemShut {NoStop}%
\bibitem [{\citenamefont {Uktamov}\ \emph {et~al.}(2025{\natexlab{c}})\citenamefont {Uktamov}, \citenamefont {Narzilloev}, \citenamefont {Hussain}, \citenamefont {Abdujabbarov},\ and\ \citenamefont {Ahmedov}}]{Uktamov:2025s2mcmc}%
  \BibitemOpen
  \bibfield  {author} {\bibinfo {author} {\bibfnamefont {U.}~\bibnamefont {Uktamov}}, \bibinfo {author} {\bibfnamefont {B.}~\bibnamefont {Narzilloev}}, \bibinfo {author} {\bibfnamefont {I.}~\bibnamefont {Hussain}}, \bibinfo {author} {\bibfnamefont {A.}~\bibnamefont {Abdujabbarov}},\ and\ \bibinfo {author} {\bibfnamefont {B.}~\bibnamefont {Ahmedov}},\ }\bibfield  {title} {\bibinfo {title} {Spinning particle motion and mcmc analysis of s2 star orbiting sgr a$^\ast$},\ }\href {https://doi.org/10.1016/j.dark.2025.102022} {\bibfield  {journal} {\bibinfo  {journal} {Physics of the Dark Universe}\ }\textbf {\bibinfo {volume} {49}},\ \bibinfo {pages} {102022} (\bibinfo {year} {2025}{\natexlab{c}})}\BibitemShut {NoStop}%
\bibitem [{\citenamefont {Uktamov}\ \emph {et~al.}(2025{\natexlab{d}})\citenamefont {Uktamov}, \citenamefont {Alloqulov}, \citenamefont {Shaymatov}, \citenamefont {Zhu},\ and\ \citenamefont {Ahmedov}}]{Uktamov:2024selfdual}%
  \BibitemOpen
  \bibfield  {author} {\bibinfo {author} {\bibfnamefont {U.}~\bibnamefont {Uktamov}}, \bibinfo {author} {\bibfnamefont {M.}~\bibnamefont {Alloqulov}}, \bibinfo {author} {\bibfnamefont {S.}~\bibnamefont {Shaymatov}}, \bibinfo {author} {\bibfnamefont {T.}~\bibnamefont {Zhu}},\ and\ \bibinfo {author} {\bibfnamefont {B.}~\bibnamefont {Ahmedov}},\ }\bibfield  {title} {\bibinfo {title} {Particle dynamics and the accretion disk around a self-dual black hole immersed in a magnetic field in loop quantum gravity},\ }\href {https://doi.org/10.1016/j.dark.2024.101743} {\bibfield  {journal} {\bibinfo  {journal} {Physics of the Dark Universe}\ }\textbf {\bibinfo {volume} {47}},\ \bibinfo {pages} {101743} (\bibinfo {year} {2025}{\natexlab{d}})},\ \Eprint {https://arxiv.org/abs/2412.01809} {arXiv:2412.01809 [gr-qc]} \BibitemShut {NoStop}%
\bibitem [{\citenamefont {Uktamov}\ \emph {et~al.}(2026{\natexlab{a}})\citenamefont {Uktamov}, \citenamefont {{\"O}vg{\"u}n}, \citenamefont {Pantig},\ and\ \citenamefont {Ahmedov}}]{UktamjonUktamov:2026dep}%
  \BibitemOpen
  \bibfield  {author} {\bibinfo {author} {\bibfnamefont {U.}~\bibnamefont {Uktamov}}, \bibinfo {author} {\bibfnamefont {A.}~\bibnamefont {{\"O}vg{\"u}n}}, \bibinfo {author} {\bibfnamefont {R.~C.}\ \bibnamefont {Pantig}},\ and\ \bibinfo {author} {\bibfnamefont {B.}~\bibnamefont {Ahmedov}},\ }\bibfield  {title} {\bibinfo {title} {{Horizon-brightened acceleration radiation and optical signatures of generic regular black holes from nonlinear electrodynamics}},\ }\href {https://doi.org/10.1140/epjc/s10052-026-15841-7} {\bibfield  {journal} {\bibinfo  {journal} {Eur. Phys. J. C}\ }\textbf {\bibinfo {volume} {86}},\ \bibinfo {pages} {631} (\bibinfo {year} {2026}{\natexlab{a}})},\ \Eprint {https://arxiv.org/abs/2602.15077} {arXiv:2602.15077 [gr-qc]} \BibitemShut {NoStop}%
\bibitem [{\citenamefont {{\"O}vg{\"u}n}\ \emph {et~al.}(2026)\citenamefont {{\"O}vg{\"u}n}, \citenamefont {Pantig}, \citenamefont {Ahmedov},\ and\ \citenamefont {Uktamov}}]{Ovgun:2025ehi}%
  \BibitemOpen
  \bibfield  {author} {\bibinfo {author} {\bibfnamefont {A.}~\bibnamefont {{\"O}vg{\"u}n}}, \bibinfo {author} {\bibfnamefont {R.~C.}\ \bibnamefont {Pantig}}, \bibinfo {author} {\bibfnamefont {B.}~\bibnamefont {Ahmedov}},\ and\ \bibinfo {author} {\bibfnamefont {U.}~\bibnamefont {Uktamov}},\ }\bibfield  {title} {\bibinfo {title} {{Acceleration radiation of freely falling atoms: Nonlinear electrodynamic effects}},\ }\href {https://doi.org/10.1016/j.dark.2026.102279} {\bibfield  {journal} {\bibinfo  {journal} {Phys. Dark Univ.}\ }\textbf {\bibinfo {volume} {52}},\ \bibinfo {pages} {102279} (\bibinfo {year} {2026})},\ \Eprint {https://arxiv.org/abs/2512.21387} {arXiv:2512.21387 [gr-qc]} \BibitemShut {NoStop}%
\bibitem [{\citenamefont {Uktamov}\ \emph {et~al.}(2026{\natexlab{b}})\citenamefont {Uktamov}, \citenamefont {Shaymatov}, \citenamefont {Ahmedov},\ and\ \citenamefont {Yuan}}]{UktamjonUktamov:2025qts}%
  \BibitemOpen
  \bibfield  {author} {\bibinfo {author} {\bibfnamefont {U.}~\bibnamefont {Uktamov}}, \bibinfo {author} {\bibfnamefont {S.}~\bibnamefont {Shaymatov}}, \bibinfo {author} {\bibfnamefont {B.}~\bibnamefont {Ahmedov}},\ and\ \bibinfo {author} {\bibfnamefont {C.}~\bibnamefont {Yuan}},\ }\bibfield  {title} {\bibinfo {title} {{New analytical model of rotating black hole with dark matter halo: constraints from EHT observations and accretion disk}},\ }\href {https://doi.org/10.1140/epjp/s13360-026-07740-3} {\bibfield  {journal} {\bibinfo  {journal} {Eur. Phys. J. Plus}\ }\textbf {\bibinfo {volume} {141}},\ \bibinfo {pages} {513} (\bibinfo {year} {2026}{\natexlab{b}})},\ \Eprint {https://arxiv.org/abs/2509.00460} {arXiv:2509.00460 [gr-qc]} \BibitemShut {NoStop}%
\bibitem [{\citenamefont {Uktamjon}\ \emph {et~al.}(2024)\citenamefont {Uktamjon}, \citenamefont {Narzilloev},\ and\ \citenamefont {Ahmedov}}]{Uktamjon:2024fjb}%
  \BibitemOpen
  \bibfield  {author} {\bibinfo {author} {\bibfnamefont {U.}~\bibnamefont {Uktamjon}}, \bibinfo {author} {\bibfnamefont {B.}~\bibnamefont {Narzilloev}},\ and\ \bibinfo {author} {\bibfnamefont {B.}~\bibnamefont {Ahmedov}},\ }\bibfield  {title} {\bibinfo {title} {{Perturbation of spacetime and electromagnetic field due to massive charged particle around electrically charged stringy black hole}},\ }\href {https://doi.org/10.1140/epjp/s13360-024-05709-8} {\bibfield  {journal} {\bibinfo  {journal} {Eur. Phys. J. Plus}\ }\textbf {\bibinfo {volume} {139}},\ \bibinfo {pages} {903} (\bibinfo {year} {2024})}\BibitemShut {NoStop}%
\bibitem [{\citenamefont {Alloqulov}\ \emph {et~al.}(2026{\natexlab{b}})\citenamefont {Alloqulov}, \citenamefont {Shaymatov}, \citenamefont {Ahmedov},\ and\ \citenamefont {Zhu}}]{Alloqulov:2026modmax}%
  \BibitemOpen
  \bibfield  {author} {\bibinfo {author} {\bibfnamefont {M.}~\bibnamefont {Alloqulov}}, \bibinfo {author} {\bibfnamefont {S.}~\bibnamefont {Shaymatov}}, \bibinfo {author} {\bibfnamefont {B.}~\bibnamefont {Ahmedov}},\ and\ \bibinfo {author} {\bibfnamefont {T.}~\bibnamefont {Zhu}},\ }\bibfield  {title} {\bibinfo {title} {Gravitational waveforms from periodic orbits around a dyonic modmax black hole},\ }\href {https://doi.org/10.1140/epjc/s10052-026-15469-7} {\bibfield  {journal} {\bibinfo  {journal} {European Physical Journal C}\ }\textbf {\bibinfo {volume} {86}},\ \bibinfo {pages} {259} (\bibinfo {year} {2026}{\natexlab{b}})},\ \Eprint {https://arxiv.org/abs/2511.15237} {arXiv:2511.15237 [gr-qc]} \BibitemShut {NoStop}%
\bibitem [{\citenamefont {Cardoso}\ \emph {et~al.}(2009)\citenamefont {Cardoso}, \citenamefont {Miranda}, \citenamefont {Berti}, \citenamefont {Witek},\ and\ \citenamefont {Zanchin}}]{Cardoso:2008bp}%
  \BibitemOpen
  \bibfield  {author} {\bibinfo {author} {\bibfnamefont {V.}~\bibnamefont {Cardoso}}, \bibinfo {author} {\bibfnamefont {A.~S.}\ \bibnamefont {Miranda}}, \bibinfo {author} {\bibfnamefont {E.}~\bibnamefont {Berti}}, \bibinfo {author} {\bibfnamefont {H.}~\bibnamefont {Witek}},\ and\ \bibinfo {author} {\bibfnamefont {V.~T.}\ \bibnamefont {Zanchin}},\ }\bibfield  {title} {\bibinfo {title} {{Geodesic stability, Lyapunov exponents and quasinormal modes}},\ }\href {https://doi.org/10.1103/PhysRevD.79.064016} {\bibfield  {journal} {\bibinfo  {journal} {Phys. Rev. D}\ }\textbf {\bibinfo {volume} {79}},\ \bibinfo {pages} {064016} (\bibinfo {year} {2009})},\ \Eprint {https://arxiv.org/abs/0812.1806} {arXiv:0812.1806 [hep-th]} \BibitemShut {NoStop}%
\bibitem [{\citenamefont {Pretorius}\ and\ \citenamefont {Khurana}(2007)}]{Pretorius:2007jn}%
  \BibitemOpen
  \bibfield  {author} {\bibinfo {author} {\bibfnamefont {F.}~\bibnamefont {Pretorius}}\ and\ \bibinfo {author} {\bibfnamefont {D.}~\bibnamefont {Khurana}},\ }\bibfield  {title} {\bibinfo {title} {Black hole mergers and unstable circular orbits},\ }\href {https://doi.org/10.1088/0264-9381/24/12/S07} {\bibfield  {journal} {\bibinfo  {journal} {Classical and Quantum Gravity}\ }\textbf {\bibinfo {volume} {24}},\ \bibinfo {pages} {S83} (\bibinfo {year} {2007})},\ \Eprint {https://arxiv.org/abs/gr-qc/0702084} {arXiv:gr-qc/0702084} \BibitemShut {NoStop}%
\bibitem [{\citenamefont {Benczik}\ \emph {et~al.}(2002)\citenamefont {Benczik}, \citenamefont {Chang}, \citenamefont {Minic}, \citenamefont {Okamura}, \citenamefont {Rayyan},\ and\ \citenamefont {Takeuchi}}]{Benczik:2002tt}%
  \BibitemOpen
  \bibfield  {author} {\bibinfo {author} {\bibfnamefont {S.}~\bibnamefont {Benczik}}, \bibinfo {author} {\bibfnamefont {L.~N.}\ \bibnamefont {Chang}}, \bibinfo {author} {\bibfnamefont {D.}~\bibnamefont {Minic}}, \bibinfo {author} {\bibfnamefont {N.}~\bibnamefont {Okamura}}, \bibinfo {author} {\bibfnamefont {S.}~\bibnamefont {Rayyan}},\ and\ \bibinfo {author} {\bibfnamefont {T.}~\bibnamefont {Takeuchi}},\ }\bibfield  {title} {\bibinfo {title} {Short distance versus long distance physics: The classical limit of the minimal length uncertainty relation},\ }\href {https://doi.org/10.1103/PhysRevD.66.026003} {\bibfield  {journal} {\bibinfo  {journal} {Physical Review D}\ }\textbf {\bibinfo {volume} {66}},\ \bibinfo {pages} {026003} (\bibinfo {year} {2002})},\ \Eprint {https://arxiv.org/abs/hep-th/0204049} {arXiv:hep-th/0204049} \BibitemShut {NoStop}%
\bibitem [{\citenamefont {Abuter}\ \emph {et~al.}(2020)\citenamefont {Abuter} \emph {et~al.}}]{GRAVITY:2020gka}%
  \BibitemOpen
  \bibfield  {author} {\bibinfo {author} {\bibfnamefont {R.}~\bibnamefont {Abuter}} \emph {et~al.} (\bibinfo {collaboration} {GRAVITY}),\ }\bibfield  {title} {\bibinfo {title} {Detection of the schwarzschild precession in the orbit of the star s2 near the galactic centre massive black hole},\ }\href {https://doi.org/10.1051/0004-6361/202037813} {\bibfield  {journal} {\bibinfo  {journal} {Astronomy \& Astrophysics}\ }\textbf {\bibinfo {volume} {636}},\ \bibinfo {pages} {L5} (\bibinfo {year} {2020})},\ \Eprint {https://arxiv.org/abs/2004.07187} {arXiv:2004.07187 [astro-ph.GA]} \BibitemShut {NoStop}%
\bibitem [{\citenamefont {Bernab{\`o}}\ \emph {et~al.}(2025)\citenamefont {Bernab{\`o}}, \citenamefont {Csizmadia}, \citenamefont {Smith}, \citenamefont {Harre}, \citenamefont {K{\'a}lm{\'a}n}, \citenamefont {Cabrera}, \citenamefont {Rauer}, \citenamefont {Gandolfi}, \citenamefont {Pino}, \citenamefont {Ehrenreich},\ and\ \citenamefont {Hatzes}}]{Bernabo:2025wasp43b}%
  \BibitemOpen
  \bibfield  {author} {\bibinfo {author} {\bibfnamefont {L.~M.}\ \bibnamefont {Bernab{\`o}}}, \bibinfo {author} {\bibfnamefont {S.}~\bibnamefont {Csizmadia}}, \bibinfo {author} {\bibfnamefont {A.~M.~S.}\ \bibnamefont {Smith}}, \bibinfo {author} {\bibfnamefont {J.-V.}\ \bibnamefont {Harre}}, \bibinfo {author} {\bibfnamefont {S.}~\bibnamefont {K{\'a}lm{\'a}n}}, \bibinfo {author} {\bibfnamefont {J.}~\bibnamefont {Cabrera}}, \bibinfo {author} {\bibfnamefont {H.}~\bibnamefont {Rauer}}, \bibinfo {author} {\bibfnamefont {D.}~\bibnamefont {Gandolfi}}, \bibinfo {author} {\bibfnamefont {L.}~\bibnamefont {Pino}}, \bibinfo {author} {\bibfnamefont {D.}~\bibnamefont {Ehrenreich}},\ and\ \bibinfo {author} {\bibfnamefont {A.}~\bibnamefont {Hatzes}},\ }\bibfield  {title} {\bibinfo {title} {Characterising wasp-43b's interior structure: Unveiling tidal decay and apsidal motion},\ }\href {https://doi.org/10.1051/0004-6361/202451994} {\bibfield  {journal} {\bibinfo  {journal} {Astronomy \& Astrophysics}\ }\textbf {\bibinfo
  {volume} {694}},\ \bibinfo {pages} {A233} (\bibinfo {year} {2025})},\ \Eprint {https://arxiv.org/abs/2501.03685} {arXiv:2501.03685 [astro-ph.EP]} \BibitemShut {NoStop}%
\bibitem [{\citenamefont {Poisson}\ and\ \citenamefont {Will}(2014)}]{PoissonWill:2014}%
  \BibitemOpen
  \bibfield  {author} {\bibinfo {author} {\bibfnamefont {E.}~\bibnamefont {Poisson}}\ and\ \bibinfo {author} {\bibfnamefont {C.~M.}\ \bibnamefont {Will}},\ }\href {https://doi.org/10.1017/CBO9781139507486} {\emph {\bibinfo {title} {Gravity: Newtonian, Post-Newtonian, Relativistic}}}\ (\bibinfo  {publisher} {Cambridge University Press},\ \bibinfo {address} {Cambridge},\ \bibinfo {year} {2014})\BibitemShut {NoStop}%
\end{thebibliography}%
\end{document}